\documentclass[english,prl, notitlepage, onecolumn, longbibliography]{revtex4-1}
\usepackage[T1]{fontenc}
\usepackage[latin9]{inputenc}
\usepackage{array}
\usepackage{float}
\usepackage{bm}
\usepackage{amsmath}
\usepackage{graphicx}

\makeatletter

\providecommand{\tabularnewline}{\\}

\usepackage{amsmath}
\usepackage{subfigure}
\usepackage{graphicx, mathtools}
\usepackage[normalem]{ulem}
\usepackage[colorlinks=true,linkcolor=MidnightBlue,urlcolor=MidnightBlue,citecolor=MidnightBlue,anchorcolor=MidnightBlue]{hyperref}\usepackage[dvipsnames]{xcolor}

\makeatother

\usepackage{babel}
\begin{document}

\title{Generalized Stokes laws for active colloids and their applications}

\author{Rajesh Singh}
\email{rsingh@imsc.res.in}

\affiliation{The Institute of Mathematical Sciences-HBNI, CIT Campus, Chennai
600113, India}

\author{R. Adhikari}
\email{rjoy@imsc.res.in, ra413@cam.ac.uk}

\affiliation{The Institute of Mathematical Sciences-HBNI, CIT Campus, Chennai
600113, India}

\affiliation{DAMTP, Centre for Mathematical Sciences, University of Cambridge,
Wilberforce Road, Cambridge CB3 0WA, UK}
\begin{abstract}
The force per unit area on the surface of a colloidal particle is
a fundamental dynamical quantity in the mechanics and statistical
mechanics of colloidal suspensions. Here we compute it in the limit
of slow viscous flow for a suspension of $N$ spherical active colloids
in which activity is represented by surface slip. Our result is best
expressed as a set of linear relations, the ``generalized Stokes laws'',
between the coefficients of a tensorial spherical harmonic expansion
of the force per unit area and the surface slip. The generalized friction
tensors in these laws are many-body functions of the colloidal configuration
and can be obtained to any desired accuracy by solving a system of
linear equations. Quantities derived from the force per unit area
- forces, torques and stresslets on the colloids and flow, pressure
and entropy production in the fluid - have succinct expressions in
terms of the generalized Stokes laws. Most notably, the active forces
and torques have a dissipative, long-ranged, many-body character that
can cause phase separation, crystallization, synchronization and a
variety of other effects observed in active suspensions. We use the
results above to derive the Langevin and Smoluchowski equations for
Brownian active suspensions, to compute active contributions to the
suspension stress and fluid pressure, and to relate the synchrony
in a lattice of harmonically trapped active colloids to entropy production.
Our results provide the basis for a microscopic theory of active Brownian
suspensions that consistently accounts for momentum conservation in
the bulk fluid and at fluid-solid boundaries.\\\\DOI: \href{http://dx.doi.org/10.1088/2399-6528/aaab0d}{10.1088/2399-6528/aaab0d}
\end{abstract}
\maketitle

\section{Introduction}

G. G. Stokes, in 1851, derived the force per unit area acting on the
surface of a sphere moving slowly in an incompressible viscous fluid
and thus obtained his celebrated laws for the drag force and torque
on a moving sphere. Einstein used these laws in his phenomenological
theory of Brownian motion and obtained a relation between the diffusion
coefficient of a spherical colloid and the friction constant in Stokes
law, the so-called Stokes-Einstein relation, the first example of
a fluctuation-dissipation relation \cite{sutherland1905,einstein1905theory,kubo1966fluctuation}.
Smoluchowski, in 1911, presented an iterative method to calculate
the force per unit area on a moving sphere in the presence of another
and thereby initiated the study of hydrodynamic interactions between
colloidal particles \cite{smoluchowski1911mutual}. His method was
refined by many authors culminating in the ``induced force'' method
of Mazur and coworkers \cite{mazur1982,ladd1988}. The theory of Brownian
motion with hydrodynamic interactions developed in parallel with major
contributions from Kirkwood \cite{kirkwood1946statistical,kirkwood1948intrinsic},
Zwanzig \cite{zwanzig1964hydrodynamic,zwanzig1969langevin}, and Batchelor
\cite{batchelor1976brownian,batchelor1977effect}, amongst others
\cite{tough1986stochastic,russel1981brownian}. Through the course
of these studies it became apparent that the force per unit area on
the surface of the colloid is the key dynamical quantity necessary
for developing both the mechanics and the statistical mechanics of
suspensions. Despite the classical nature of the subject, the study
of colloidal suspensions remains a research area of sustained fecundity
\cite{dhont1996introduction}. 

Bearing testimony to this is the recent spate of experimental work
on colloidal particles that produce flow in the ambient fluid in the
absence of externally applied forces or torques \cite{paxton2004,howse2007self,jiang2010active,ebbens2010pursuit,palacci2013living}.
The stress in the fluid produced by this flow reacts back on the colloids
and may cause them to translate or rotate. The hydrodynamic interactions
between such colloids are without analogue in their classical counterparts.
An enormously rich variety of phenomena have been observed in suspensions
of such ``active colloids'' and a systematic study of this new class
of colloidal particles is now underway. It has also been realized
that phoretic phenomena in gradients of externally applied fields
\cite{anderson1989colloid}, the swimming of micro-organisms \cite{brennen1977,jahn1972locomotion},
and the self-propulsion of drops \cite{thutupalli2011swarming,niu2017self},
though distinct in the microscopic mechanisms that produce fluid flow,
share many points of similarity with synthetic active colloids when
viewed at the suspension scale. Their study has been revitalized \cite{drescher2009,drescher2010,petroff2015fast}
and subsumed into the field of active colloids. 

The principal feature common to both synthetic active colloids and
classical phoretic motion in external fields is the appearance of
surface forces, of a range much shorter than the colloidal dimensions,
that drive fluid flow in a thin region at the colloid-fluid boundary.
The typical width of this boundary layer is several nanometers, which
is approximately a thousand times smaller than the size of the colloidal
particle. This wide separation of scales makes it possible to partition
the problem of determining the global fluid flow into an interior
problem inside the boundary layer, where the flow is driven by the
surface forces, and an exterior problem outside the boundary layer,
where the exterior fluid is presented with a boundary condition determined
by the flow in the boundary layer. The non-universal aspects of the
problem are contained, therefore, entirely within the boundary layer
leaving the exterior fluid flow to assume universal forms determined
by the boundary conditions. Smoluchowski was the first to pursue this
approach in his study of electrophoresis \cite{smoluchowski1903}.
It was used by Derjaguin to study diffusiophoresis \cite{derjaguin1947kinetic}
and other kinds of phoretic motion \cite{derjaguin1987surface} and
subsequent applications have confirmed the generality and the power
of this approach \cite{anderson1989colloid}. 

The swimming of ciliated microorganisms presents another instance
of the separation of scales, as the length of the cilium on the surface
of such organisms is typically a hundred times smaller than the dimensions
of their body. The flow in the ciliary layer can be obtained from
the prescribed kinematics of each cilium and the exterior flow then
obtained by matching boundary conditions at the interface. This analysis
has been developed extensively by Blake \cite{blake1971a}, following
Lighthill's abstract analysis of the squirming motion of a sphere
\cite{lighthill1952}. These authors only considered axisymmetric
squirming motion and it is only recently that the most general non-axisymmetric
motion has been studied systematically \cite{anderson1991,ghose2014irreducible,pak2014generalized,felderhof1994small,felderhof2016stokesian,pedley2016spherical}. 

The self-propulsion of drops is distinct from the previous examples
due to the permeability of the interface that separates the internal
and external fluids. In these systems, motion is caused by expulsion
and suction of fluid at different parts of the interface while approximately
maintaining the volume of the drop. In contrast to the previous examples,
where the slip flow is tangential to the boundary, in this case, the
slip flow has a normal component and thus belongs to the class of
osmophoretic phenomena \cite{anderson1983movement,anderson1989colloid}.
The method presented here applies such motion, provided that the interfacial
forces are sufficiently strong to maintain the spherical drop shape
and the rate of dissolution is sufficiently slow that the volume remains
approximately constant on the time scales of drop motion. 

In all of the above examples, the matching condition at the edge of
the boundary layer is the continuity of the fluid velocity, which
now contains, in addition to the rigid body motion of the colloid,
the contribution from the active flow $\boldsymbol{v}^{\mathcal{A}}$.
This ``slip'' velocity is a general, possibly time-dependent, vector
field on the colloid surface, subject only to the constraint of mass
conservation $\int\boldsymbol{v}^{\mathcal{A}}\cdot\mathbf{n}\,dS=0$.
The fluid mechanical problem is solved when the exterior flow is obtained
in terms of the slip. The force per unit area on the colloidal surface
follows directly from the Cauchy stress in the fluid and the motion
of the colloid is obtained, when colloidal inertia is negligible,
by setting the net force and net torque to zero. 

The correct expression for the force per unit area was first obtained
by Blake \cite{blake1971a} for the special case of axisymmetric slip
on an isolated sphere distant from all boundaries. This expression
was used to compute the work done on the fluid by the slip. However,
extensions of this analysis to (i) the most general form of slip in
(ii) a suspension of many spheres and (iii) including the effect of
proximate boundaries remain to be completed. The availability of such
an expression would immediately allow the mechanics and statistical
mechanics of active colloidal suspensions to reach the level of development
of their passive counterparts. We address ourselves to this task here.

Below, we obtain the force per unit area in a suspension of $N$ spherical
active colloids with the most general form of slip and including the
effect of proximate boundaries such as plane walls or plane interfaces.
Our results are best expressed as an infinite set of linear relations
between the irreducible coefficients of a tensorial harmonic expansion
of the force per unit area and the slip. As these contain the Stokes
laws for the force and torque as special cases, we refer to them as
``generalized Stokes laws''. 

The generalization has several aspects. First, expressions are obtained
for every irreducible mode of the force per unit area, and not just
the first two modes representing the force and the torque. Second,
the many-body nature of the force per unit area in a suspension of
$N$ colloids is explicitly considered, whereas Stokes laws in their
original form apply only to a single particle. Third, the effect of
proximate boundaries, such as plane walls, are taken into account
to obtain corrections of the kind first derived by Faxén \cite{faxen1922widerstand}
to the original form of Stokes laws. The generalized Stokes laws contain
new coefficients, analogous to the translational and rotational friction
tensors, that are many-body functions of the colloidal configuration.
We show how these generalized friction tensors can be calculated,
to any desired degree of accuracy, from the solution of a linear system.
This material is presented in sections \ref{sec:generalized-stokes}
and \ref{sec:Boundary-integral-solution}.

The generalized Stokes laws immediately provide the quantities of
interest to suspension mechanics: the force, the torque, and the stresslet
on the colloids. Through their use, succinct expressions are obtained
for the fluid flow, the fluid pressure, and the entropy production.
We apply these results in the following sections to demonstrate their
utility. In section \ref{sec:langevin-description} we use results
for the force and torque to obtain the over-damped Langevin equations
for hydrodynamically interacting Brownian active colloids. The corresponding
Smoluchowski equation is also derived and the activity-induced breakdown
of the fluctuation-dissipation relation is demonstrated. A contracted
description for the one-body distribution function, the first member
of a BBGKY-like hierarchy, is derived and the active contribution
is highlighted. In section \ref{sec:Suspension-stress}, we use the
result for the stresslet to obtain an exact expression for the Landau-Lifshitz-Batchelor
suspension stress as a function of colloidal configuration. We confirm
that the active contribution to the suspension stress vanishes to
linear order in volume fraction in an isotropic suspension. In contrast,
the active contributions for a non-isotropic suspension of contractile
(or extensile) active colloids leads to an increase (or decrease)
of the suspension stress. In section \ref{sec:Active-pressure} we
turn to mechanical quantities in the fluid, focusing on the fluid
pressure in a suspension of active colloids confined to the interior
of a spherical volume. We find that the dynamics of the colloids and
the distribution of the fluid pressure are different for suspensions
of extensile and contractile colloids. The difference in the collective
dynamics of the colloids is understood qualitatively from the fluid
flow created by the active colloids. In section \ref{sec:Dynamics-optical},
we use the Langevin equations to study the collective motion of active
colloids in harmonic traps. The traps are centered on a lattice and
we consider only one active colloid per trap. Through detailed numerical
simulations, we find that a one-dimensional lattice of traps supports
time-independent configurations but a two-dimensional lattice of traps
only supports time-dependent configurations. In particular, we show
that the colloids exhibit a synchronized dynamics about the axis of
symmetry in a square lattice of traps and we relate synchrony to the
entropy production in a triangular lattice. We conclude in section
\ref{sec:Discussion-and-summary} by discussing other methods for
studying active suspensions and by suggesting further applications
of the generalized Stokes laws.

The results presented here are based on several keys ideas from our
previous work on the suspensions of active colloids. The tensorial
harmonic expansion of the slip was introduced in \cite{ghose2014irreducible}
and the infinite-dimensional linear system that results from a Galerkin
discretization of the boundary integral equation in this expansion
basis was obtained in \cite{singh2015many}. The linear system was
only partially solved to obtain the rigid body motion directly in
terms of propulsion tensors. In particular, it was not recognized
that a complete solution of the linear system for the coefficients
of the force per unit area provide \emph{all} quantities necessary
for developing a complete theory of the mechanics and statistical
mechanics of active suspensions. It is this aspect of the problem
that is explored here.

The methods of this paper have been applied previously to rationalize
the stability of, and transitions between, spontaneously aggregated
states of active colloids near boundaries in terms of the hydrodynamically
mediated active forces and torques \cite{singh2016crystallization,thutupalli2017boundaries}.
Such a dynamical point of view, in which forces and torques are primary
and velocities and angular velocities are derived, yields a generic
theory for flow-induced phase separation (FIPS) which is complementary
to the kinematical theory of motility-induced phase separation (MIPS)
\cite{tailleur2008statistical,cates2010arrested,cates2013active,cates2015,henkes2011active,redner2013structure}.
We foresee many other instances where FIPS should provide an accurate
representation of the physical forces that drive active aggregation. 

\section{generalized Stokes laws\label{sec:generalized-stokes}}

\begin{figure*}[t]
\centering\includegraphics[width=0.92\textwidth]{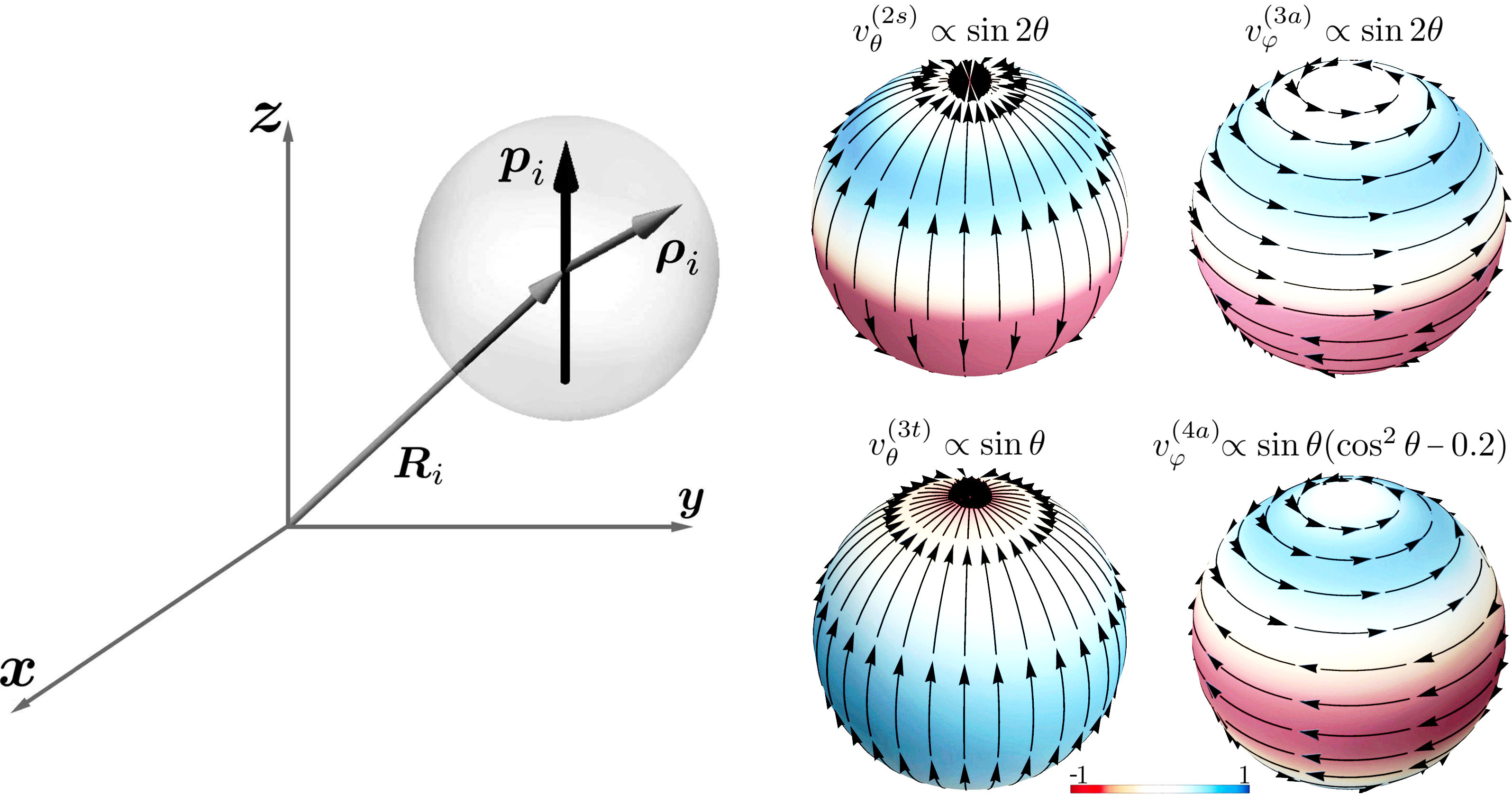}\caption{Coordinate system to describe spherical active colloids (left panel).
The center of mass of $i$th colloid is at ${\boldsymbol{R}}_{i}$
and a unit vector ${\boldsymbol{p}}_{i}$ describes its orientation
while $\bm{\rho}_{i}$ is the radius vector. The plots on the right
are the streamlines of the surface velocity, overlaid on the pseudo-color
plot of the logarithm of the normalized flow speed, due to four leading
slip modes as described in the text about Eq.(\ref{eq:-slip-truncate}).
The modes are parameterized uniaxially in terms of $\boldsymbol{p}$,
which is chosen to be along $\hat{\boldsymbol{z}}$ direction, without
losing generality, such that $\boldsymbol{p}=\cos\theta\,\boldsymbol{\hat{\rho}}-\sin\theta\,\boldsymbol{\hat{\theta}}$,
where $(\boldsymbol{\hat{\rho}},\,\boldsymbol{\hat{\theta}},\,\boldsymbol{\boldsymbol{\hat{\varphi}}})$
are the spherical polar unit vectors. \label{fig:Coordinate-system-used}}
\end{figure*}

We now present our main result relating the force per unit area $\boldsymbol{f}$
to the active slip $\boldsymbol{v}_{i}^{\mathcal{A}}$ for $i=1,\ldots,N$
active colloidal spheres of radius $b$ suspended in an incompressible
viscous fluid of viscosity $\eta$. The $i$-th sphere centered at
$\boldsymbol{R}_{i}$ has radius vector $\boldsymbol{\rho}_{i}$ and
orientation vector $\boldsymbol{p}_{i}$ as shown in Fig.(\ref{fig:Coordinate-system-used}).
The velocity and angular velocity of the sphere are $\mathbf{V}_{i}$
and $\mathbf{\Omega}_{i}$ respectively. In the microhydrodynamic
regime, Newton's equations reduce to instantaneous balance of forces
and torques
\begin{equation}
\mathbf{F}_{i}^{H}+\mathbf{F}_{i}^{P}+\hat{\mathbf{F}}_{i}^{\mathcal{}}=0,\quad\mathbf{T}_{i}^{H}+\mathbf{T}_{i}^{P}+\hat{\mathbf{T}}_{i}=0.\label{eq:newtons}
\end{equation}
Here $\mathbf{F}_{i}^{H}=\int\boldsymbol{f}\,d\text{S}_{i}$ are hydrodynamic
forces, $\mathbf{F}_{i}^{P}$ are body forces and $\hat{\mathbf{F}}_{i}^{\mathcal{}}$
are Brownian forces, while $\mathbf{T}_{i}^{H}=\int\boldsymbol{\rho}_{i}\mathbf{\times}\boldsymbol{f}\,d\text{S}_{i}$,
$\mathbf{T}_{i}^{P}$ and $\hat{\mathbf{T}}_{i}$ are corresponding
torques. The velocity $\boldsymbol{u}(\boldsymbol{R}_{i}+\boldsymbol{\rho}_{i})$
of the fluid at a point $\boldsymbol{R}_{i}+\boldsymbol{\rho}_{i}$
on the colloid boundary is a sum of its rigid body motion and the
active slip $\boldsymbol{v}_{i}^{\mathcal{A}}$
\begin{equation}
\boldsymbol{u}(\boldsymbol{R}_{i}+\boldsymbol{\rho}_{i})=\mathbf{V}_{i}+\bm{\Omega}_{i}\times\bm{\rho}_{i}+\boldsymbol{v}_{i}^{\mathcal{A}}(\bm{\rho}_{i}).\label{eq:slip-RBM-BC}
\end{equation}
The wide applicability of slip to model motion driven by interfacial
forces has been mentioned above. To these, we add the recent observation
that $\boldsymbol{v}_{i}^{\mathcal{A}}$ need not be identified with
the material surface of the colloid but can, instead, be thought of
as the velocity at the surface of an effective sphere enclosing a
non-spherical active body such as the biflagellate green algae \emph{C.
Reinhardti} \cite{ghose2014irreducible}. A sphere with slip, therefore,
provides a good representation for a surprisingly large variety of
active flows and should be preferred as the simplest dynamical model
of a finite-sized active body. 

Both the slip and the force per unit area are vector fields on the
sphere, which makes harmonic expansions natural. We choose the tensorial
spherical harmonics $\mathbf{Y}^{(l)}(\bm{\hat{\rho}})=(-1)^{l}\rho^{l+1}\bm{\nabla}^{(l)}\rho^{-1}$,
where $\bm{\nabla}^{(l)}=\bm{\nabla}_{\alpha_{1}}\dots\bm{\nabla}_{\alpha_{l}}$,
as the expansion basis for reasons discussed below. These harmonics
are dimensionless, symmetric, irreducible Cartesian tensors of rank
$l$ that form a complete, orthogonal basis on the sphere \cite{weinert1980spherical,applequist1989traceless,hess2015tensors}.
The expansion of the slip and the force per unit area on the surface
of $i$th colloid is
\begin{alignat}{1}
\boldsymbol{v}_{i}^{\mathcal{A}}\big(\boldsymbol{R}_{i}+\bm{\rho}_{i}\big)=\sum_{l=1}^{\infty}\frac{1}{(l-1)!(2l-3)!!}\,\mathbf{V}_{i}^{(l)}\cdot\mathbf{Y}^{(l-1)}(\bm{\hat{\rho}}_{i}),\qquad\boldsymbol{f}\big(\boldsymbol{R}_{i}+\bm{\rho}_{i}\big)= & \sum_{l=1}^{\infty}\frac{2l-1}{4\pi b^{2}}\,\mathbf{F}_{i}^{(l)}\cdot\mathbf{Y}^{(l-1)}(\bm{\hat{\rho}}_{i}).\label{eq:boundaryFields-Yl}
\end{alignat}
The expansion coefficients $\mathbf{V}_{i}^{(l)}$ and $\mathbf{F}_{i}^{(l)}$,
of dimensions of force and velocity respectively, are tensors of rank
$l$, symmetric and irreducible in their last $l-1$ indices \cite{singh2015many}.
Elementary group theory assures \cite{tung1985group} that they can
be expressed as the sum of three irreducible tensors of rank $l$,
$l-1$ and $l-2$. The three irreducible parts are
\begin{eqnarray}
\mathbf{V}_{i}^{(l\sigma)}=\mathbf{P}^{(l\sigma)}\cdot\mathbf{V}_{i}^{(l)},\qquad\mathbf{F}_{i}^{(l\sigma)}=\mathbf{P}^{(l\sigma)}\cdot\mathbf{F}_{i}^{(l)},\qquad\label{eq:project-vlsigma}
\end{eqnarray}
where the index $\sigma=s,\,a\text{ and }t$, labels the symmetric
irreducible, antisymmetric and trace parts of the reducible tensors.
The operator $\mathbf{P}^{(ls)}=\boldsymbol{\Delta}^{(l)}$ extracts
the symmetric irreducible part, $\mathbf{P}^{(la)}=\boldsymbol{\Delta}^{(l-1)}\boldsymbol{\boldsymbol{\varepsilon}}$
the antisymmetric part and $\mathbf{P}^{(lt)}=\boldsymbol{\delta}$
the trace of the operand. Here $\mathbf{\Delta}^{(l)}$ is tensor
of rank $2l$, projecting any $l$th order tensor to its symmetric
irreducible form \cite{hess2015tensors}, $\boldsymbol{\varepsilon}$
is the Levi-Civita tensor and $\boldsymbol{\mathbf{\delta}}$ is the
Kronecker delta. As the $\mathbf{V}_{i}^{(l\sigma)}$ are irreducible
tensors, it is natural to parametrize them in terms of the tensorial
spherical harmonics. Their \emph{uniaxial} parametrizations are
\begin{align}
\mathbf{V}_{i}^{(ls)}=V_{0}^{(ls)}\,\mathbf{Y}^{(l)}(\boldsymbol{p}_{i}),\qquad\mathbf{V}_{i}^{(la)}=V_{0}^{(la)}\,\mathbf{Y}^{(l-1)}(\boldsymbol{p}_{i}), & \qquad\mathbf{V}_{i}^{(lt)}=V_{0}^{(lt)}\,\mathbf{Y}^{(l-2)}(\boldsymbol{p}_{i}).\label{eq:uniaxial-parametrization}
\end{align}
It follows that the coefficients are either even (apolar) or odd (polar)
under inversion symmetry $\boldsymbol{p}_{i}\rightarrow-\boldsymbol{p}_{i}$.
The leading terms of the expansion, categorized according to their
symmetry under inversion and mirror reflection, are \begin{subequations}\label{eq:-slip-truncate}
\begin{eqnarray}
\boldsymbol{v}_{i}^{\mathcal{A}}(\bm{\rho}_{i})=-\underbrace{\mathbf{V}_{i}^{\mathcal{A}}+\tfrac{1}{15}\mathbf{V}_{i}^{(3t)}\hspace{-0.12cm}\cdot\mathbf{Y}^{(2)}(\bm{\hat{\rho}}_{i})}_{\mathrm{achiral,\,polar}}+\underbrace{\mathbf{V}_{i}^{(2s)}\hspace{-0.12cm}\cdot\mathbf{Y}^{(1)}(\bm{\hat{\rho}}_{i})}_{\mathrm{achiral,\,apolar}}-\underbrace{\mathbf{\Omega}_{i}^{\mathcal{A}}\times\bm{\rho}_{i}-\tfrac{1}{60}\boldsymbol{\varepsilon}\cdot\mathbf{V}_{i}^{(4a)}\hspace{-0.12cm}\cdot\mathbf{Y}^{(3)}(\bm{\hat{\rho}}_{i})}_{\mathrm{chiral,\,polar}}-\underbrace{\tfrac{1}{9}\boldsymbol{\varepsilon}\cdot\mathbf{V}_{i}^{(3a)}\hspace{-0.12cm}\cdot\mathbf{Y}^{(2)}(\bm{\hat{\rho}}_{i})}_{\mathrm{chiral,\thinspace apolar}}\label{eq:slip-truncation-1}
\end{eqnarray}
\begin{alignat}{1}
4\pi a^{2}\,\mathbf{V}_{i}^{\mathcal{A}}=-\int\boldsymbol{v}_{i}^{\mathcal{A}}(\bm{\rho}_{i})dS_{i},\qquad4\pi a^{2}\,\bm{\Omega}_{i}^{\mathcal{A}}=-\frac{3}{2a^{2}}\int\bm{\rho}_{i}\times\boldsymbol{v}_{i}^{\mathcal{A}}(\bm{\rho}_{i})dS_{i},\label{eq:one-body}
\end{alignat}
\end{subequations}Here $\mathbf{V}_{i}^{\mathcal{A}}\equiv-\mathbf{V}_{i}^{(1s)}$
is the active translational velocity, while $\mathbf{\Omega}_{i}^{\mathcal{A}}\equiv-\tfrac{1}{b}\mathbf{V}_{i}^{(2a)}$
is the active angular velocity for a sphere in unbounded medium \cite{anderson1991,stone1996,ghose2014irreducible}.
The tangential flows generated by these leading terms are shown in
Fig.(\ref{fig:Coordinate-system-used}). Their complete form, including
radial terms, is provided in Table \ref{tab:Slip-velocity-on} of
the Appendix. 

The tensorial harmonics have several advantages over the more common
vector spherical harmonics or surface polynomials \cite{zick1982stokes,ichiki2002improvement}.
First, both basis functions and expansion coefficients transform as
Cartesian tensors which allows physical quantities like the force,
the torque and the stresslet to be represented covariantly. Second,
the linear independence of the basis functions ensures that linear
systems for the coefficients are always of full rank. Third, the basis
functions are well-suited for Taylor expansions \cite{singh2015many}
and their addition theorems are considerably simpler than the usual
vector spherical harmonics. This property allows for fast methods
of summing long-ranged harmonics \cite{shanker2007accelerated} that
are less complex than the classical methods \cite{greengard1987fast}.
Previous studies of spherical passive suspensions have amply demonstrated
some of these advantages \cite{brunn1976effect,mazur1982,ladd1988,ghose2014irreducible,singh2015many}.
The problem, now, is to relate the unknown values of the traction
coefficients $\mathbf{F}_{i}^{(l\sigma)}$ to the known values of
the slip coefficients $\mathbf{V}_{i}^{(l\sigma)}$.

In the regime of slow viscous flow, the fluid velocity $\boldsymbol{u}$
in Eq.(\ref{eq:slip-RBM-BC}) obeys the Stokes equation. The linearity
of the governing equation and of the boundary condition implies, then,
a linear relation between the velocity and traction coefficients
\begin{eqnarray}
 &  & \mathbf{F}_{i}^{(l\sigma)}=\underbrace{-\boldsymbol{\gamma}_{ij}^{(l\sigma,\,1s)}\cdot\mathbf{V}_{j}-\boldsymbol{\gamma}_{ij}^{(l\sigma,\,2a)}\cdot\mathbf{\Omega}_{j}}_{\text{rigid body motion}}-\sum_{l'\sigma'=1s}^{\infty}\underbrace{\boldsymbol{\gamma}_{ij}^{(l\sigma,\,l'\sigma')}\cdot\mathbf{V}_{j}^{(l'\sigma')}}_{\text{activity}},\label{eq:traction-l-sigma}
\end{eqnarray}
where repeated particle indices are summed over. We call the above
infinite set the \textit{generalized Stokes laws.} Their best-known
special cases are Stokes laws for translation $\mathbf{F}_{i}=-6\pi\eta b\mathbf{V}_{i}$
and rotation $\mathbf{T}_{i}=-8\pi\eta b^{3}\boldsymbol{\Omega}_{i}$.
The first two terms are passive frictional contributions due to rigid
body motion while the remaining terms are active contributions due
to slip. The $\boldsymbol{\gamma}_{ij}^{(l\sigma,\,l'\sigma')}$ are
generalized friction tensors and their symmetry properties, to which
we shall return below, allow them to be interpreted as generalized
Onsager coefficients. Their rank varies from $l+l'$ to $l+l^{\prime}-4$
depending on the values of $\sigma$ and $\sigma^{\prime}$. The friction
tensors depend on the positions of \emph{all }spheres, reflecting
the many-body character of the hydrodynamic interaction but, due to
isotropy, are independent of the sphere orientations. The traction
coefficients, however, depend on the orientations of the spheres carried
in the coefficients of the slip. Given the velocity at the boundary
of each sphere, the friction tensors completely determine the force
per unit area on every sphere. In the next section, we derive exact
expressions for the generalized friction tensors in terms of Green's
functions of Stokes flow.

\section{Boundary integral solution\label{sec:Boundary-integral-solution}}

The most direct way of computing the generalized friction tensors
is through the boundary integral representation of Stokes flow \cite{fkg1930bandwertaufgaben,ladyzhenskaya1969,youngren1975stokes,zick1982stokes,pozrikidis1992,muldowney1995spectral,cheng2005heritage,singh2015many}\begin{subequations}\label{eq:BIE}
\begin{alignat}{1}
u_{\alpha}(\boldsymbol{r})=-\int G_{\alpha\beta}(\boldsymbol{r},\,\boldsymbol{r}_{j})\,f_{\beta}(\boldsymbol{r}_{j})\,d\text{S}_{j} & +\int K_{\beta\alpha\gamma}(\boldsymbol{r},\,\boldsymbol{r}_{j})\hat{\rho}_{\gamma}u_{\beta}(\boldsymbol{r}_{j})\,d\text{S}_{j},\label{eq:BIE-v}\\
p(\boldsymbol{r})=-\int P_{\beta}(\boldsymbol{r},\,\boldsymbol{r}_{j})\,f_{\beta}(\boldsymbol{r}_{j})\,d\text{S}_{j} & +\int Q_{\alpha\beta}(\boldsymbol{r},\,\boldsymbol{r}_{j})\hat{\rho}_{\alpha}u_{\beta}(\boldsymbol{r}_{j})\,d\text{S}_{j}.\label{eq:BIE-1}
\end{alignat}
\end{subequations}where $p$ is the fluid pressure. In the above,
repeated Cartesian and particle indices are summed over and points
on the boundary of the sphere are given by ${\bf r}_{i}=\boldsymbol{R}_{i}+\bm{\rho}_{i}$.
The kernels in the boundary integral representations are the Green's
function $\mathbf{G}$, the pressure vector $\mathbf{P}$, its gradient
$\mathbf{Q}=2\eta\boldsymbol{\nabla}\mathbf{P}$, and the stress tensor
$\mathbf{K}$. They satisfy\begin{subequations}
\begin{alignat}{1}
\nabla_{\alpha}G_{\alpha\beta}(\boldsymbol{r},\,\boldsymbol{r'})=0,\qquad-\nabla_{\alpha}P_{\beta}(\boldsymbol{r},\,\boldsymbol{r'})+\eta\nabla^{2}G_{\alpha\beta}(\boldsymbol{r},\,\boldsymbol{r'})=-\delta\left(\boldsymbol{r}-\boldsymbol{r'}\right)\delta_{\alpha\beta},\\
K_{\alpha\beta\gamma}(\boldsymbol{r},\boldsymbol{r'})=-\delta_{\alpha\gamma}P_{\beta}(\boldsymbol{r},\,\boldsymbol{r'})+\eta\left(\nabla_{\gamma}G_{\alpha\beta}(\boldsymbol{r},\,\boldsymbol{r'})+\nabla_{\alpha}G_{\beta\gamma}(\boldsymbol{r},\,\boldsymbol{r'})\right).\quad\,\,
\end{alignat}
\end{subequations}The flow and pressure satisfying the Stokes equation,
$\nabla\cdot\boldsymbol{u}=0$ and $\nabla\cdot\boldsymbol{\sigma}$
= 0 with $\bm{\sigma}=-p\boldsymbol{I}+\eta(\bm{\nabla}\boldsymbol{u}+(\bm{\nabla}\boldsymbol{u})^{T})$
the Cauchy stress in a fluid of viscosity $\eta$, is the sum of the
``single-layer'' integral of the local force per unit area $\boldsymbol{f}=\text{\ensuremath{\hat{\boldsymbol{\rho}}}}_{i}\cdot\boldsymbol{\sigma}$
and the ``double-layer'' integral of the boundary velocity $\mathbf{V}_{i}+\bm{\Omega}_{i}\times\bm{\rho}_{i}+\boldsymbol{v}_{i}^{\mathcal{A}}(\bm{\rho}_{i}).$
From now on, we shall refer to $\boldsymbol{f}$ as the traction.
Enforcing the boundary conditions in the integral representation produces
a Fredholm integral equation of the first kind for the unknown force
per unit area. The problem is thereby reduced from the solution of
a partial differential equation in a three-dimensional volume to the
solution of an integral equation over two-dimensional surfaces. 

Several methods of solution have been developed for Fredholm integral
equations for Stokes flows \cite{youngren1975stokes,zick1982stokes,ladd1988,pozrikidis1992,muldowney1995spectral,kim2005,singh2015many}.
These differ in their choice of formulation, discretization, and strategy
of minimizing the residual. Here, we use a direct formulation, in
which the single-layer density is the physical force per unit area,
in contrast to other formulations where such an interpretation is
unavailable \cite{zick1982stokes,ladd1988}. The choice of discretization
in terms of the tensorial spherical harmonics is natural for spheres
and such global basis functions yield the greatest accuracy for the
least number of unknowns \cite{muldowney1995spectral}. Finally, Galerkin's
method of minimizing the residual is chosen as it yields a self-adjoint
linear system for the coefficients, an advantageous property for numerical
solutions. This choice of formulation, discretization and residual
minimization is of greatest utility for spheres, as the matrix elements
of the linear system can be evaluated analytically \cite{singh2015many}
and the numerical quadrature typically associated with the Galerkin
method can be avoided entirely. We note that for particles of non-spherical
shape, these advantages are no longer available. 

The linear system we obtain using this direct formulation with the
global tensorial harmonic basis functions and Galerkin's method of
minimizing the residual is \cite{singh2015many}
\begin{alignat}{1}
 & -\boldsymbol{G}_{ij}^{(l,\,l')}(\boldsymbol{R}_{i},\boldsymbol{R}_{j})\cdot\mathbf{F}_{j}^{(l')}+\boldsymbol{K}_{ij}^{(l,\,l')}(\boldsymbol{R}_{i},\boldsymbol{R}_{j})\cdot\mathbf{V}_{j}^{(l)}=\begin{cases}
\tfrac{1}{2}\left(\mathbf{V}_{i}-\mathbf{V}_{i}^{\mathcal{A}}\right)\qquad & l\sigma=1s,\\
\\
\tfrac{1}{2}\left(b\mathbf{\Omega}_{i}-b\mathbf{\Omega}_{i}^{\mathcal{A}}\right)\quad\qquad & l\sigma=2a,\\
\\
\tfrac{1}{2}\mathbf{V}_{i}^{(l\sigma)} & \text{otherwise,}
\end{cases}\label{eq:linear-system}
\end{alignat}
where the matrix elements $\boldsymbol{G}_{ij}^{(l,\,l')}(\boldsymbol{R}_{i},\boldsymbol{R}_{j})$
and $\boldsymbol{K}_{ij}^{(l,\,l')}(\boldsymbol{R}_{i},\boldsymbol{R}_{j})$
\begin{alignat*}{1}
\boldsymbol{G}_{ij}^{(l,\,l')}(\boldsymbol{R}_{i},\boldsymbol{R}_{j}) & =\frac{(2l-1)(2l'-1)}{(4\pi b^{2})^{2}}\int\mathbf{Y}^{(l-1)}(\hat{\bm{\rho}}_{i})\mathbf{G}(\boldsymbol{R}_{i}+\bm{\rho}_{i},\boldsymbol{R}_{j}+\bm{\rho}_{j})\mathbf{Y}^{(l'-1)}(\hat{\bm{\rho}}_{j})\,d\text{S}_{i}d\text{S}_{j},\\
\boldsymbol{K}_{ij}^{(l,\,l')}(\boldsymbol{R}_{i},\boldsymbol{R}_{j}) & =\frac{(2l-1)}{4\pi b^{2}(l'-1)!(2l'-3)!!}\int\mathbf{Y}^{(l-1)}(\hat{\bm{\rho}}_{i})\mathbf{K}(\boldsymbol{R}_{i}+\bm{\rho}_{i},\boldsymbol{R}_{j}+\bm{\rho}_{j})\cdot\bm{\hat{\rho}}_{j}\mathbf{Y}^{(l'-1)}(\hat{\bm{\rho}}_{j})\,d\text{S}_{i}d\text{S}_{j},
\end{alignat*}
are given in terms of the Green's function $\mathbf{G}(\boldsymbol{R}_{i},\boldsymbol{R}_{j})$
and its derivatives, see Appendix \ref{appendix:boundary-integrals}.
This linear system differs in important ways from superficially similar
linear systems derived for passive colloids \cite{ladd1988}. First,
the right-hand sides of the equations corresponding to the rigid body
motions contain self-propulsion and self-rotation contributions; these
are absent in passive colloids. Second, the double-layer integral
adds a non-trivial contribution to the linear system; this is trivial
in passive colloids as rigid body motions are in the eigenspace of
the double-layer. Finally, the equations for the non-rigid body modes
are inhomogeneous, both due to double layer and slip contributions;
these equations are homogeneous for passive systems as both double-layer
and slip contributions are zero. Linear systems for passive colloids,
derived earlier by Mazur \cite{mazur1982} and by Ladd \cite{ladd1988},
are recovered from the above when all $\mathbf{V}_{i}^{(l\sigma)}$
are set to zero. 

The generalized Stokes laws are obtained as solutions of this linear
system. The formal solution of the discrete linear system yields the
following expression for the friction tensors
\begin{equation}
\boldsymbol{\gamma}^{(l\sigma,\,l'\sigma')}=\mathbf{P}^{(l\sigma)}\cdot\left[\boldsymbol{G}^{-1}\cdot\left(\tfrac{1}{2}\boldsymbol{I}-\boldsymbol{K}\right)\right]^{(l,\,l')}\cdot\mathbf{P}^{(l'\sigma')}.\label{eq:generalized-friction-tensor}
\end{equation}
In the above, we have introduced the block matrices $\boldsymbol{G}$
and $\boldsymbol{K}$ whose $(l,\,l^{\prime})$ element in the $ij$
block are $\boldsymbol{G}_{ij}^{(l,\,l')}(\boldsymbol{R}_{i},\boldsymbol{R}_{j})$
and $\boldsymbol{K}_{ij}^{(l,\,l')}(\boldsymbol{R}_{i},\boldsymbol{R}_{j})$
respectively. The many-body character of the friction tensors is apparent
from this expressions, as the inverse of $\boldsymbol{G}$ cannot
be expressed in terms of sums of the pairwise interactions encoded
in its matrix elements. Since the matrix elements only involve particle
positions and not orientations, the friction tensors are, likewise,
independent of particle orientation. 

The method of solution for the linear system, or equivalently, of
inverting $\boldsymbol{G}$ must be chosen according to need. For
analytical solutions, Jacobi's iterative method is straightforward
and its widespread use in Stokes flows can be traced back to Smoluchowski's
method of reflections \cite{ichiki2001many}. Other tractable analytical
methods rely on series expansion of $\boldsymbol{G}^{-1}$ in powers
of $\mathbf{G}$ \cite{mazur1985hydrodynamic}. We give explicit analytical
expressions for the friction tensors obtained from the Jacobi method
in Appendix \ref{appendix:iterative-sol}. For numerical solutions,
iterative solvers with faster rates of convergence must be used. Typically,
these search for the inverse in the Krylov subspace of $\boldsymbol{G}$
and have the advantage of requiring, instead of $\boldsymbol{G}$
, only its action on a vector of the appropriate size. The need to
store a large dense matrix is thereby avoided. For the self-adjoint
linear system above, the stable, efficient and accurate conjugate
gradient method \cite{hestenes1952methods} may be used. This requires
$O(M^{2})$ computational effort for $M$ unknowns when matrix-vector
products are computed directly. The use of fast summation methods
can reduce the cost to $O(M\log M)$ \cite{barnes1986hierarchical,sierou2001accelerated}
or even $O(M)$ \cite{greengard1987fast,sangani1996n,ladd1994numericala}.
With these preliminary remarks, we postpone the study of numerical
solutions for the friction tensors to a future work.

Expressions for the friction tensors in terms of Green's functions
of the Stokes equation have been derived in Appendix \ref{appendix:boundary-integrals}.
Examples of the Green's functions of Stokes flow for typical flow
geometries used in the experiments of active colloids is provided
in Table \ref{tab:geometries-G}. As shown in Appendix \ref{appendix:boundary-integrals},
the leading order contribution to the friction tensor is
\begin{equation}
\boldsymbol{\gamma}_{ij}^{(l\sigma,\,l^{\prime}\sigma^{\prime})}\sim\nabla_{i}^{l-1}\nabla_{j}^{l^{\prime}-1}\mathbf{G}(\boldsymbol{R}_{i},\boldsymbol{R}_{j}).
\end{equation}
In an unbounded fluid, this decays as $|\boldsymbol{R}_{i}-\boldsymbol{R}_{j}|^{-(l+l^{\prime}-1)}$
and is thus long-ranged for $l+l^{\prime}\leq4.$ Thus, all slip modes
up to $l=3$ produce long-ranged forces and up to $l=2$ produce long-ranged
torques. Notably, the active forces and torques have contributions
even without self-propulsion, $\mathbf{V}_{i}^{\mathcal{A}}=0$, or
self-rotation, $\mathsf{\mathbf{\Omega}}_{j}^{\mathcal{A}}=0$, indicating
that colloids can be ``active'' without necessarily being self-propelling
or self-rotating. The friction tensors depend on the positions of
all the particles and are thus many-body functions of the instantaneous
colloidal configurations. A remarkable feature of active forces and
torques is that they depend not only on the relative position of the
particles, as in a passive suspension of spheres, but also through
the slip moments, on their orientations. Thus active colloids, even
if they are geometrically isotropic, are hydrodynamically anisotropic.
An intuitive understanding of the orientation-dependent forces and
torques can be obtained by studying Fig.(\ref{fig:orientation-dependence})
of the Appendix, where a change in orientation of the active colloid
alters the flow field it ``carries'' and consequently the stresses
that are transmitted by this flow field. Finally, though these forces
and torques appear in Newton's equations, they are, emphatically,
not body forces and torques: they are the sum of the dissipative surface
forces that act at the fluid-solid boundary. 

The boundary integral representation of Eq.(\ref{eq:BIE}) can be
used to obtain explicit expressions for the fluid velocity and the
fluid pressure at any point in the bulk of the fluid. By expanding
the traction and velocity in the boundary integral expression we obtain
expression for fluid flow and pressure in terms of the tensorial coefficients
as\begin{subequations}\label{eq:v-p}
\begin{alignat}{1}
\boldsymbol{u}(\boldsymbol{r}) & =\sum_{l=1}^{\infty}\Big(-\boldsymbol{G}_{j}^{(l)}(\boldsymbol{r},\boldsymbol{R}_{j})\cdot\mathbf{F}_{j}^{(l)}+\boldsymbol{K}_{j}^{(l)}(\boldsymbol{r},\boldsymbol{R}_{j})\cdot\mathbf{V}_{j}^{(l)}\Big),\label{eq:vel-BI}\\
p(\boldsymbol{r}) & =\sum_{l=1}^{\infty}\Big(-\boldsymbol{P}_{j}^{(l)}(\boldsymbol{r},\boldsymbol{R}_{j})\cdot\mathbf{F}_{j}^{(l)}+\boldsymbol{Q}_{j}^{(l)}(\boldsymbol{r},\boldsymbol{R}_{j})\cdot\mathbf{V}_{j}^{(l)}\Big),\label{eq:press-BI}
\end{alignat}
\end{subequations}where the boundary integrals for fluid flow and
pressure are\begin{subequations}\label{eq:bi-v-p}
\begin{alignat}{1}
\boldsymbol{G}_{j}^{(l)}(\boldsymbol{r},\,\boldsymbol{R}_{j}) & =\frac{2l-1}{4\pi b^{2}}\int\mathbf{G}(\boldsymbol{r},\,\boldsymbol{R}_{j}+\bm{\rho}_{j})\mathbf{Y}^{(l-1)}(\hat{\bm{\rho}}_{j})d\text{S}_{j},\\
\boldsymbol{K}_{j}^{(l)}(\boldsymbol{r},\,\boldsymbol{R}_{j}) & =\frac{1}{(l-1)!(2l-3)!!}\int\mathbf{K}(\boldsymbol{r},\,\boldsymbol{R}_{j}+\bm{\rho}_{j})\cdot\bm{\hat{\rho}}_{j}\mathbf{Y}^{(l-1)}(\hat{\bm{\rho}}_{j})d\text{S}_{j},\\
\boldsymbol{P}_{j}^{(l)}(\boldsymbol{r},\,\boldsymbol{R}_{j}) & =\frac{2l-1}{4\pi b^{2}}\int\mathbf{P}(\boldsymbol{r},\,\boldsymbol{R}_{j}+\bm{\rho}_{j})\mathbf{Y}^{(l-1)}(\hat{\bm{\rho}}_{j})\,d\text{S}_{j},\\
\boldsymbol{Q}_{j}^{(l)}(\boldsymbol{r},\,\boldsymbol{R}_{j}) & =\frac{1}{(l-1)!(2l-3)!!}\int\mathbf{Q}(\boldsymbol{r},\,\boldsymbol{R}_{j}+\bm{\rho}_{j})\cdot\bm{\hat{\rho}}_{j}\mathbf{Y}^{(l-1)}(\hat{\bm{\rho}}_{j})\,d\text{S}_{j}.
\end{alignat}
\end{subequations}These boundary integrals are obtained explicitly
in terms of the Green's function of Stokes flow and fluid pressure
and their derivatives. These results are derived in Appendix \ref{appendix:fluid-vel-pressure}.
Using the generalized Stokes laws of Eq.(\ref{eq:traction-l-sigma})
and balance conditions of Eq.(\ref{eq:newtons}), the unknown traction
coefficients can be written in terms of the known slip coefficients
and the body forces and torques. The expression for fluid velocity
and pressure can be then expressed entirely in terms of known quantities
as\begin{subequations}\label{eq:p-v-knowns}
\begin{alignat}{1}
\boldsymbol{u}(\boldsymbol{r})=\, & \boldsymbol{G}_{j}^{(1s)}(\boldsymbol{r},\,\boldsymbol{R}_{j})\cdot\mathbf{F}_{j}^{P}+\boldsymbol{G}_{j}^{(2a)}(\boldsymbol{r},\,\boldsymbol{R}_{j})\cdot\mathbf{T}_{j}^{P}+\sum_{l\sigma=2s}^{\infty}\boldsymbol{\Pi}_{j}^{(l\sigma)}\cdot\mathbf{V}_{j}^{(l\sigma)},\label{eq:fluid-flow-irred}\\
p(\boldsymbol{r})=\, & \boldsymbol{P}_{j}^{(1)}(\boldsymbol{r},\,\boldsymbol{R}_{j})\cdot\mathbf{F}_{j}^{P}+\sum_{l=2}^{\infty}\boldsymbol{\Lambda}_{j}^{(ls)}\cdot\mathbf{V}_{j}^{(ls)}.\label{eq:fuid-pressure-irred}
\end{alignat}
\end{subequations}Here $\boldsymbol{G}_{i}^{(1s)}$ and $\boldsymbol{G}_{i}^{(2a)}$
are matrices which, respectively, relate the fluid velocity with body
forces and torques while $\boldsymbol{\Pi}_{i}^{(l\sigma)}$ give
the linear relation between coefficients of slip velocity to fluid
velocity. The contribution of body forces to pressure is given by
$\boldsymbol{P}_{i}^{(1s)}$, while $\boldsymbol{\Lambda}_{i}^{(ls)}$
gives the contributions from the symmetric irreducible modes of the
active slip. It is instructive to note that only symmetric traceless
parts $\mathbf{V}^{(ls)}$, of the slip modes contribute to the fluid
pressure, as pressure is harmonic, unlike the fluid velocity which
is biharmonic. $\boldsymbol{\Pi}_{i}^{(l\sigma)}$ and $\boldsymbol{\Lambda}_{i}^{(l\sigma)}$
are tensors of rank $l$ and $l+1$, respectively, and in an unbounded
fluid, decay as $r^{-l}$ and $r^{-l+1}$ with distance $r$. 

The power dissipation in the volume of the fluid \cite{landau1959fluid}
can be reduced to integrals on the colloidal boundaries using the
divergence theorem
\begin{eqnarray}
\dot{\mathcal{E}} & =\int\boldsymbol{\sigma}:(\nabla\boldsymbol{u})\,dV= & -\sum_{i}^{N}\int\boldsymbol{f}(\boldsymbol{R}_{i}+\mathbf{\boldsymbol{\rho}}_{i})\cdot\boldsymbol{u}(\boldsymbol{R}_{i}+\mathbf{\boldsymbol{\rho}}_{i})\,d\text{S}_{i},
\end{eqnarray}
and using Eq.(\ref{eq:boundaryFields-Yl}) along with the orthogonality
of the basis, can be expressed in terms of the expansion coefficients
as
\begin{alignat}{1}
\dot{\mathcal{E}} & =-\sum_{i}^{N}\left(\mathbf{F}_{i}^{H}\cdot\mathbf{V}_{i}+\mathbf{T}_{i}^{H}\cdot\mathbf{\Omega}_{i}+\sum_{l\sigma=1s}^{\infty}\mathbf{F}_{i}^{(l\sigma)}\cdot\mathbf{V}_{i}^{(l\sigma)}\right).
\end{alignat}
The generalized Stokes laws can now be used to eliminate the unknown
traction coefficients for the known slip coefficients to obtain
\begin{eqnarray}
\dot{\mathcal{E}} & = & -\mathbf{F}_{i}^{H}\cdot\mathbf{V}_{i}-\mathbf{T}_{i}^{H}\cdot\mathbf{\Omega}_{i}+\mathbf{V}_{i}^{(l\sigma)}\cdot\boldsymbol{\gamma}_{ij}^{(l\sigma,\,1s)}\cdot\mathbf{V}_{j}+\mathbf{V}_{i}^{(l\sigma)}\cdot\boldsymbol{\gamma}_{ij}^{(l\sigma,\,2a)}\cdot\mathbf{\Omega}_{j}+\mathbf{V}_{i}^{(l\sigma)}\cdot\boldsymbol{\gamma}_{ij}^{(l\sigma,\,l'\sigma')}\cdot\mathbf{V}_{j}^{(l'\sigma')}.
\end{eqnarray}
The positivity of power dissipation requires the friction tensors
to be positive-definite. Proofs of the latter are available in \cite{wajnryb2013generalization,singh2017fluctuation}.

The differences with passive colloids are noteworthy: the colloids
can produce flow in the absence of external forces and torques and
both this flow and the pressure depend on the position and the orientation
of the colloids. The latter fact is illustrated in Fig.(\ref{fig:orientation-dependence}).
The contributions from the slip makes flow in active suspensions intrinsically
more rich and leads to novel phenomena with no analogue in passive
suspensions.

To summarize, the main result of this section is Eq.(\ref{eq:generalized-friction-tensor}),
which gives exact expressions for the generalized friction tensors
in terms of the matrix elements of the discretized boundary integral
equation. An approximate solution to this, in leading powers of distance
between the colloids, is provided in Appendix \ref{appendix:iterative-sol}.
We now turn to applications of the above results.

\section{Langevin and Smoluchowski descriptions \label{sec:langevin-description}}

The net hydrodynamic force $\mathbf{F}_{i}^{H}$ and torque $\mathbf{T}_{i}^{H}$
acting on the \textit{i}-th colloid are related to its first two irreducible
coefficients of the traction $\boldsymbol{f}$ and from the generalized
Stokes laws these are\begin{subequations}\label{force-formulation}
\begin{alignat}{1}
\mathbf{F}_{i}^{H}= & -\boldsymbol{\gamma}_{ij}^{TT}\mathbf{\cdot V}_{j}-\boldsymbol{\gamma}_{ij}^{TR}\mathbf{\cdot\boldsymbol{\Omega}}_{j}-\sum_{l\sigma=1s}^{\infty}\boldsymbol{\gamma}_{ij}^{(T,\,l\sigma)}\cdot\mathbf{V}_{j}^{(l\sigma)},\label{eq:linear-force-torque}\\
\mathbf{T}_{i}^{H}= & -\boldsymbol{\gamma}_{ij}^{RT}\mathbf{\cdot V}_{j}-\boldsymbol{\gamma}_{ij}^{RR}\mathbf{\cdot\boldsymbol{\Omega}}_{j}-\sum_{l\sigma=1s}^{\infty}\boldsymbol{\gamma}_{ij}^{(R,\,l\sigma)}\cdot\mathbf{V}_{j}^{(l\sigma)}.\label{eq:torque-expression}
\end{alignat}
\end{subequations}Here, the $\boldsymbol{\gamma}_{ij}^{\alpha\beta}$
with $\alpha,\beta=T,R$ are relabellings of the four friction tensors
$\boldsymbol{\gamma}_{ij}^{(l\sigma,l'\sigma')}$ with $l\sigma,l'\sigma'=1s,2a$,
where the correspondences are $T\leftrightarrow1s$ and $R\leftrightarrow2a$.
The $\boldsymbol{\gamma}_{ij}^{(T,l\sigma)}$ and $\boldsymbol{\gamma}_{ij}^{(R,l\sigma)}$
are relabellings of the friction tensors $\boldsymbol{\gamma}_{ij}^{(1s,l\sigma)}$
and $\boldsymbol{\gamma}_{ij}^{(2a,l\sigma)}$ respectively. The first
two terms in the force and torque expressions above are the usual
rigid body drag \cite{mazur1982,ladd1988} while the remaining terms
are the contributions from the slip. The effect of activity is then
clear: it adds long-ranged, many-body correlated, orientation-dependent
dissipative forces and torques to the familiar Stokes drags.

The hydrodynamic forces and torques obtained above can be used to
construct the Langevin equations describing the motion of active colloids
in a thermally fluctuating fluid \cite{hauge1973fluctuating,fox1970contributions,bedeaux1974brownian,beenakker1983b,singh2017fluctuation,roux1992brownian,zwanzig1964hydrodynamic}.
This is obtained from the balance of hydrodynamic, body and Brownian
forces and torques as provided in Eq.(\ref{eq:newtons}). The Brownian
forces and torques, $\hat{\mathbf{F}}_{i}$ and $\mathbf{\hat{T}}_{i}$,
are zero-mean, Gaussian white noises and the fluctuation-dissipation
relation fixes their variances to be\begin{subequations}
\begin{align}
\langle\hat{\mathbf{F}}_{i}\rangle & =0,\qquad\langle\hat{\mathbf{F}}_{i}(t)\,\hat{\mathbf{F}}_{j}(t')\rangle=2k_{B}T\,\boldsymbol{\gamma}_{ij}^{TT}\delta(t-t'),\qquad\langle\hat{\mathbf{F}}_{i}(t)\,\hat{\mathbf{T}}_{j}(t')\rangle=2k_{B}T\,\boldsymbol{\gamma}_{ij}^{TR}\delta(t-t'),\\
\langle\hat{\mathbf{T}}_{i}\rangle & =0,\qquad\langle\hat{\mathbf{T}}_{i}(t)\,\hat{\mathbf{F}}_{j}(t')\rangle=2k_{B}T\,\boldsymbol{\gamma}_{ij}^{RT}\delta(t-t'),\qquad\langle\hat{\mathbf{T}}_{i}(t)\,\hat{\mathbf{T}}_{j}(t')\rangle=2k_{B}T\,\boldsymbol{\gamma}_{ij}^{RR}\delta(t-t').
\end{align}
\end{subequations}where $k_{B}$ is the Boltzmann constant and $T$
is the temperature. We do not consider any fluctuations corresponding
to activity, since its inherently non-equilibrium nature precludes
any possible balance between fluctuation and dissipation. Using explicit
forms, the Langevin equations for active colloids with hydrodynamic
interactions are\begin{subequations}\label{force-formulation-1}
\begin{alignat}{1}
-\boldsymbol{\gamma}_{ij}^{TT}\mathbf{\cdot V}_{j}-\boldsymbol{\gamma}_{ij}^{TR}\mathbf{\cdot\,\boldsymbol{\Omega}}_{j}+\mathbf{F}_{i}^{P}+\hat{\mathbf{F}}_{i}-\sum_{l\sigma=1s}^{\infty}\boldsymbol{\gamma}_{ij}^{(T,\,l\sigma)}\cdot\mathbf{V}_{j}^{(l\sigma)}=0,\label{eq:linear-force-torque-1}\\
-\boldsymbol{\gamma}_{ij}^{RT}\mathbf{\cdot V}_{j}-\boldsymbol{\gamma}_{ij}^{RR}\mathbf{\cdot\,\boldsymbol{\Omega}}_{j}+\mathbf{T}_{i}^{P}+\hat{\mathbf{T}}_{i}-\sum_{l\sigma=1s}^{\infty}\boldsymbol{\gamma}_{ij}^{(R,\,l\sigma)}\cdot\mathbf{V}_{j}^{(l\sigma)}=0.\label{eq:torque-expression-1}
\end{alignat}
\end{subequations}

The above equations contain forces due to Stokes drags, body forces,
thermal fluctuations, and activity. Motion is driven by the last three
terms and their relative importance can be captured by two ratios.
We choose the first of these to be the ratio of active and body forces
and the second to be the ratio of thermal and active forces. Similar
considerations apply for the torque balance. These motivate the introduction
of the following dimensionless numbers
\begin{equation}
\mathcal{A}_{T}=\frac{\big|\boldsymbol{\gamma}_{ij}^{(T,\,l\sigma)}\cdot\mathbf{V}_{j}^{(l\sigma)}\big|}{\big|\mathbf{F}_{i}^{P}\big|},\quad\mathcal{A}_{R}=\frac{\big|\boldsymbol{\gamma}_{ij}^{(R,\,l\sigma)}\cdot\mathbf{V}_{j}^{(l\sigma)}\big|}{\big|\mathbf{T}_{i}^{P}\big|},\qquad\mathcal{B}_{T}=\frac{\big|\hat{\mathbf{F}}_{i}\big|}{\big|\boldsymbol{\gamma}_{ij}^{(T,\,l\sigma)}\cdot\mathbf{V}_{j}^{(l\sigma)}\big|},\quad\mathcal{B}_{R}=\frac{\big|\hat{\mathbf{T}}_{i}\big|}{\big|\boldsymbol{\gamma}_{ij}^{(R,\,l\sigma)}\cdot\mathbf{V}_{j}^{(l\sigma)}\big|}.\label{eq:dimls-number}
\end{equation}
Here $\mathcal{A}_{T}$ and $\mathcal{A}_{R}$ are ``activity'' numbers
quantifying the relative importance of active and body terms \cite{jayaraman2012,laskar2013,laskar2015brownian}
while $\mathcal{B}_{T}$ and $\mathcal{B}_{R}$ are ``Brown'' numbers
quantifying the relative importance of thermal and active terms. We
estimate these numbers for two typical active colloidal systems below. 

Explicit Langevin equations for the velocity and angular velocity
are obtained by inverting Eq.(\ref{force-formulation-1}). Since the
$\boldsymbol{\gamma}_{ij}^{\alpha\beta}$ and the $\boldsymbol{\gamma}_{ij}^{(\alpha,l\sigma)}$are
identical for $l\sigma=1s,2a$, it is convenient to group the velocity
with the self-propulsion and the angular velocity with the self-rotation
so that the summation is from $l\sigma=2s$ onward. With this regrouping,
the result is\begin{subequations}\label{eq:mobility-formulation}
\begin{alignat}{1}
\mathsf{\mathbf{V}}_{i} & =\boldsymbol{\mu}_{ij}^{TT}\cdot\mathbf{F}_{j}^{P}+\boldsymbol{\mu}_{ij}^{TR}\cdot\mathbf{T}_{j}^{P}+\sqrt{2k_{B}T\bm{\mu}_{ij}^{TT}}\cdot\bm{\eta}_{j}^{T}+\sqrt{2k_{B}T\bm{\mu}_{ij}^{TR}}\cdot\bm{\zeta}_{j}^{R}+\sum_{l\sigma=2s}^{\infty}\boldsymbol{\pi}_{ij}^{(T,\,l\sigma)}\cdot\mathbf{\mathsf{\mathbf{V}}}_{j}^{(l\sigma)}+\mathsf{\mathbf{V}}_{i}^{\mathcal{A}},\label{eq:RBM-velocity}\\
\mathsf{\mathbf{\Omega}}_{i} & =\underbrace{\boldsymbol{\mu}_{ij}^{RT}\cdot\mathbf{F}_{j}^{P}+\boldsymbol{\mu}_{ij}^{RR}\cdot\mathbf{T}_{j}^{P}}_{\mathrm{Passive}}+\underbrace{\sqrt{2k_{B}T\bm{\mu}_{ij}^{RT}}\cdot\bm{\bm{\zeta}}_{j}^{T}+\sqrt{2k_{B}T\bm{\mu}_{ij}^{RR}}\cdot\bm{\eta}_{j}^{R}}_{\mathrm{Brownian}}+\sum_{l\sigma=2s}^{\infty}\underbrace{\boldsymbol{\pi}_{ij}^{(R,\,l\sigma)}\cdot\mathbf{\mathsf{\mathbf{V}}}_{j}^{(l\sigma)}+\mathsf{\mathbf{\Omega}}_{i}^{\mathcal{A}}}_{\mathrm{Active}}.\label{eq:RBM-angular-velocity}
\end{alignat}
\end{subequations}Here $\boldsymbol{\eta}^{\alpha}$, $\boldsymbol{\zeta}^{\alpha}$are
Gaussian white noises with zero-mean and variances $1$, $1/b$ respectively.
The mobility matrices $\boldsymbol{\mu}_{ij}^{\alpha\beta}$ are inverses
of the friction matrices $\boldsymbol{\gamma}_{ij}^{\alpha\beta}$
\cite{happel1965low,felderhof1977hydrodynamic,mazur1982,schmitz1982mobility,nunan1984effective,ladd1988,durlofsky1987dynamic,brady1988dynamic,kim2005,cichocki1994friction}.
The propulsion tensors $\boldsymbol{\pi}_{ij}^{(\alpha,\,l\sigma)}$,
first introduced in \cite{singh2015many}, relate the rigid body motion
to modes of the active velocity. They are related to the slip friction
tensors introduced here by\begin{subequations}\label{eq:pi-mu-relation}
\begin{alignat}{1}
-\bm{\pi}_{ij}^{(T,\,l\sigma)} & =\boldsymbol{\mu}_{ik}^{TT}\cdot\boldsymbol{\gamma}_{kj}^{(T,\,l\sigma)}+\boldsymbol{\mu}_{ik}^{TR}\cdot\boldsymbol{\gamma}_{kj}^{(R,\,l\sigma)},\\
-\bm{\pi}_{ij}^{(R,\,l\sigma)} & =\boldsymbol{\mu}_{ik}^{RT}\cdot\boldsymbol{\gamma}_{kj}^{(T,\,l\sigma)}+\boldsymbol{\mu}_{ik}^{RR}\cdot\boldsymbol{\gamma}_{kj}^{(R,\,l\sigma)}.
\end{alignat}
\end{subequations} The translational propulsion tensors $\bm{\pi}_{ij}^{(\text{T},\,l\sigma)}$
are dimensionless while the rotational propulsion tensors $\bm{\pi}_{ij}^{(R,\,l\sigma)}$
have dimensions of inverse length. The above form of the propulsion
tensors is particularly useful when mobilities are evaluated by combining
the far-field and near-field lubrication contributions. Through this
approximation, the need to resolve the rapidly varying flow at particle
contact can be avoided, and accurate results can be obtained by keeping
only the long-ranged contributions when solving the linear system. 

Stochastic trajectories can be obtained by integrating the kinematic
equations
\begin{equation}
\dot{\boldsymbol{R}}_{i}=\mathbf{V}_{i},\quad\dot{\boldsymbol{p}}_{i}=\boldsymbol{\Omega}_{i}\times\boldsymbol{p}_{i}.
\end{equation}
using the standard Brownian dynamics integrators, for example that
due to Ermak and McCammon \cite{ermak1978}. We note that the explicit
form of the Langevin equation was first obtained in \cite{laskar2015brownian}
using heuristic arguments.

The Smoluchowski equation for the distribution function $\Psi(\boldsymbol{R}_{1},\dots,\,\boldsymbol{R}_{N};\,\boldsymbol{p}_{1},\dots,\,\boldsymbol{p}_{N})\equiv\Psi(\boldsymbol{R}^{N};\,\boldsymbol{p}^{N})$
of positions and orientations follows immediately from the Langevin
equations. We write it in the form of a conservation law in configuration
space
\begin{align}
\frac{\partial\Psi}{\partial t} & =-\sum_{ij}\boldsymbol{\mathcal{L}}_{ij}\Psi=-\sum_{ij}\left(\boldsymbol{\nabla}_{{\scriptscriptstyle \boldsymbol{R}_{i}}}\cdot\boldsymbol{\mathcal{V}}_{{\scriptscriptstyle \boldsymbol{R}}_{ij}}+\boldsymbol{p}_{i}\times\boldsymbol{\nabla}_{{\scriptscriptstyle \boldsymbol{p}_{i}}}\cdot\boldsymbol{\mathcal{V}}_{{\scriptscriptstyle \boldsymbol{p}}_{ij}}\right)\Psi,\label{eq:dist-func}
\end{align}
where the ``velocities'' $\boldsymbol{\mathcal{V}}_{{\scriptscriptstyle \boldsymbol{R}}_{ij}}$
and $\boldsymbol{\mathcal{V}}_{{\scriptscriptstyle \boldsymbol{p}}_{ij}}$
are
\begin{align*}
\boldsymbol{\mathcal{V}}_{{\scriptscriptstyle \boldsymbol{R}}_{ij}} & =\boldsymbol{\mu}_{ij}^{TT}\cdot\left(\mathbf{F}_{j}^{P}-k_{B}T\,\boldsymbol{\nabla}{}_{{\scriptscriptstyle \boldsymbol{R}_{j}}}\right)+\boldsymbol{\mu}_{ij}^{TR}\cdot\left(\mathbf{T}_{j}^{P}-k_{B}T\,\boldsymbol{p}_{j}\times\boldsymbol{\nabla}_{{\scriptscriptstyle \boldsymbol{p}}_{j}}\right)+\sum_{l\sigma=2s}^{\infty}\boldsymbol{\pi}_{ij}^{(T,\,l\sigma)}\cdot\mathbf{V}_{j}^{(l\sigma)}+\mathbf{V}_{i}^{\mathcal{A}},\\
\boldsymbol{\mathcal{V}}_{{\scriptscriptstyle \boldsymbol{p}}_{ij}} & =\boldsymbol{\mu}_{ij}^{RT}\cdot\left(\mathbf{F}_{j}^{P}-k_{B}T\,\boldsymbol{\nabla}_{{\scriptscriptstyle \boldsymbol{R}_{j}}}\right)+\boldsymbol{\mu}_{ij}^{RR}\cdot\left(\mathbf{T}_{j}^{P}-k_{B}T\,\boldsymbol{p}_{j}\times\boldsymbol{\nabla}_{{\scriptscriptstyle \boldsymbol{p}}_{j}}\right)+\sum_{l\sigma=2s}^{\infty}\boldsymbol{\pi}_{ij}^{(R,\,l\sigma)}\cdot\mathbf{V}_{j}^{(l\sigma)}+\mathbf{\Omega}_{i}^{\mathcal{A}}.
\end{align*}
Here $\mathbf{F}_{j}^{P}=-\mathbf{\boldsymbol{\nabla}}_{{\scriptscriptstyle \boldsymbol{R}}_{j}}U$,
$\mathbf{T}_{j}^{P}=-\boldsymbol{p}_{j}\times\boldsymbol{\nabla}_{{\scriptscriptstyle \boldsymbol{p}}_{j}}U$,
and $U$ is a potential that contains both positional and orientational
interactions. In the absence of activity, the equation obeys the fluctuation-dissipation
relation and the Gibbs distribution $\Psi\sim\exp(-U/k_{B}T$) is
the stationary solution. This solution has a vanishing current. In
contrast, in the presence of activity, the Gibbs distribution is no
longer the stationary distribution and in general, only the divergence
of the current is zero. Circulation currents are a generic possibility
in such circumstances and have, indeed, been observed in models of
active colloidal systems \cite{nash2010,fodor2016far,tailleur2008statistical,singh2015many}. 

The Smoluchowski equation above is a partial differential equation
in $6N$ variables and is generally intractable as it stands. However,
by standard reduction methods, equations for the one and two particle
distribution functions may be obtained from it, providing a principled
way of recovering coarse-grained macroscopic equations from the microscopic
dynamics \cite{altenberger1973light,phillies1974effects,ackerson1976correlations,felderhof1978diffusion,ohtsuki1982diffusion}.
To this end, we define the $n$-body density $c^{(n)}$ as,
\begin{equation}
c^{(n)}(\boldsymbol{R}^{n},\,\boldsymbol{p}^{n},\,t)=\frac{N!}{(N-n)!}\int\Psi(\boldsymbol{R}^{N};\,\boldsymbol{p}^{N})\prod_{i=n+1}^{N}d\boldsymbol{R}_{i}d\boldsymbol{p}_{i}.\,\label{eq:n-particle-density}
\end{equation}
The dynamics of one-body density is then obtained by using the Smoluchowski
equation
\begin{align*}
\frac{\partial c^{(1)}}{\partial t}=- & N\sum_{ij}\int\boldsymbol{\mathcal{L}}_{ij}\Psi(\boldsymbol{R}^{N};\,\boldsymbol{p}^{N})\,\prod_{i=2}^{N}d\boldsymbol{R}_{i}d\boldsymbol{p}_{i}=-N\sum_{ij}\left(\boldsymbol{\nabla}_{{\scriptscriptstyle \boldsymbol{R}_{1}}}\cdot\int\boldsymbol{\mathcal{V}}_{{\scriptscriptstyle \boldsymbol{R}}_{ij}}\Psi+\boldsymbol{p}_{i}\times\boldsymbol{\nabla}_{{\scriptscriptstyle \boldsymbol{R}_{1}}}\cdot\int\boldsymbol{\mathcal{V}}_{{\scriptscriptstyle \boldsymbol{p}}_{ij}}\Psi\right)\prod_{i=2}^{N}d\boldsymbol{R}_{i}d\boldsymbol{p}_{i}.
\end{align*}
We now use the fact that one-body mobilities are diagonal and scalar
$\boldsymbol{\mu}_{11}^{RR}=\mu^{R}\,\boldsymbol{I}$, $\boldsymbol{\mu}_{11}^{TT}=\mu^{T}\,\boldsymbol{I}$
in an unbounded fluid flow. The one-body density, then satisfies,
\begin{align*}
\frac{\partial c^{(1)}}{\partial t} & =\boldsymbol{\nabla}_{{\scriptscriptstyle \boldsymbol{R}_{1}}}\cdot\left(\mu^{T}\boldsymbol{\nabla}_{{\scriptscriptstyle \boldsymbol{R}_{1}}}U-\mathbf{V}_{1}^{\mathcal{A}}\right)c^{(1)}+k_{B}T\,\mu^{T}\,\boldsymbol{\nabla}_{{\scriptscriptstyle \boldsymbol{R}_{1}}}^{2}c^{(1)}+\mu^{R}\,\boldsymbol{p}_{1}\times\boldsymbol{\nabla}_{{\scriptscriptstyle \boldsymbol{p}}_{1}}\cdot(c^{(1)}\boldsymbol{p}_{1}\times\boldsymbol{\nabla}_{{\scriptscriptstyle \boldsymbol{p}}_{1}}U-\mathbf{\Omega}_{1}^{\mathcal{A}}+k_{B}T\,\boldsymbol{p}_{1}\times\boldsymbol{\nabla}_{{\scriptscriptstyle \boldsymbol{p}}_{1}}c^{(1)})\\
 & -N(N-1)\left(\boldsymbol{\nabla}_{{\scriptscriptstyle \boldsymbol{R}_{1}}}\cdot\int\boldsymbol{\mathcal{V}}_{{\scriptscriptstyle \boldsymbol{R}}_{12}}\Psi\,d\boldsymbol{R}_{2}d\boldsymbol{p}_{2}+\boldsymbol{p}_{1}\times\boldsymbol{\nabla}_{{\scriptscriptstyle \boldsymbol{R}_{1}}}\cdot\int\boldsymbol{\mathcal{V}}_{{\scriptscriptstyle \boldsymbol{p}}_{12}}\Psi\,d\boldsymbol{R}_{2}d\boldsymbol{p}_{2}\right)\prod_{i=3}^{N}d\boldsymbol{R}_{i}d\boldsymbol{p}_{i}.
\end{align*}
We now use the definition of $c^{(2)}$ and the operator $\boldsymbol{\mathcal{L}}_{ij}$
to obtain
\begin{align}
\frac{\partial c^{(1)}}{\partial t} & =-\boldsymbol{\mathcal{L}}_{11}\,c^{(1)}-\int\boldsymbol{\mathcal{L}}_{12}\,c^{(2)}\,d\boldsymbol{R}_{2}d\boldsymbol{p}_{2}.\label{eq:one-particle-density}
\end{align}
In obtaining the above expression we have assumed that the potentials
generating the forces and torques have no inter-particle contributions.
From the above equation, it is obvious that we need $c^{(2)}$ to
obtain dynamics of $c^{(1)}$. It can be shown that $c^{(2)}$ is
obtained in terms of $c^{(3)}$ and so on till we come to $N-$colloid
density function. This system of equations is called the Bogoliubov-Born-Green-Kirkwood-Yvon
(BBGKY) hierarchy and closure schemes to these equations in available
\cite{resibois1977classical}. The coarse-grained description, thus
obtained from the microscopic equations, can be used to derive the
hydrodynamic description of active matter \cite{ramaswamy2010,marchetti2013}.
Another possible application of the distribution function is the study
of spinodal decomposition \cite{dhont1996introduction} in active
colloidal suspensions. We will pursue these directions of investigations
in detail elsewhere. We note that a subset of the above reduced description
has been obtained previously in \cite{menzel2016dynamical}. 

It is instructive to estimate the relative magnitude of the active
and Brownian terms in typical experiments. An active particle moving
at a speed $v_{s}$ has a typical active force $F_{\mathcal{\mathcal{A}}}\sim6\pi\eta bv_{s}$
acting on it. Similarly the torque acting on a colloid due to active
spin on its axis at an angular speed $\omega_{s}$ is $T_{\mathcal{A}}\sim8\pi\eta b^{3}\omega_{s}$.
We estimate the typical active force and torque for the experiment
in \cite{petroff2015fast}, where the radius of the colloid is $b=4\,\mu\text{m}$,
$v_{s}=500\,\mu\text{m/s}$ and fluid viscosity $\eta=10^{-3}\,\text{kg/ms}$.
The active force is then $F_{\mathcal{\mathcal{A}}}\sim40\times10^{-12}\,\text{N}$,
while the typical Brownian forces are of order $\mathcal{O}\left(k_{B}\text{T/b}\right)\sim10^{-15}\,\text{N}$.
This implies that the dimensionless Brown number $\mathcal{B}_{T}\sim10^{-4}$
is very small. For the same experiment, the angular speed of the colloids
is $\omega_{s}\sim50\,s^{-1}$, which implies typical active torque
is of order $T_{\mathcal{\mathcal{A}}}\sim10^{-16}$Nm. The Brownian
torques are of order $\mathcal{O}\left(k_{B}\text{T}\right)\sim10^{-21}\,\text{N}$,
which implies that the rotational Brown number $\mathcal{B}_{R}$
is of the order of $10^{-5}$. The radius of the green algae in \cite{guasto2010}
is $\sim3\mu m$ and it swimming speed is $134\mu/s$. The Brown number
for this experiment is then $\mathcal{B}^{T}\sim10^{-3}$. In another
set of experiment on bacteria \cite{chen2015dynamic,sokolov2007concentration}
and Janus colloids \cite{jiang2010active,palacci2013living}, the
size $b\sim1\,\mu\text{m}$, and the speed $v_{s}\sim10\,\mu\text{m/s}$.
This leads to $F_{\mathcal{\mathcal{A}}}\sim10^{-13}\,\text{N}$,
which implies that the Brown number is $\mathcal{B}_{T}\sim10^{-2}$.
Thus Brown numbers $\mathcal{B}_{T},\,\mathcal{B}_{R}\rightarrow0$
for commonly studied active colloids. 

\section{Suspension stress\label{sec:Suspension-stress}}

Landau and Lifshitz showed that the stress $\boldsymbol{\Sigma}^{{H}}$
in a suspension of force-free particles, averaged over scales large
compared to the particle size, is given by $\boldsymbol{\Sigma}^{{H}}=2\eta\boldsymbol{E}+\boldsymbol{\Sigma}^{P},$
where $\boldsymbol{E}$ is the macroscopic strain rate in a suspension
of volume $V$ and
\begin{eqnarray}
 & \boldsymbol{\Sigma}^{P} & =\frac{1}{V}\sum_{i}\int\left[\boldsymbol{f}^{H}(\boldsymbol{R}_{i}+\boldsymbol{\rho}_{i})\,\bm{\rho}_{i}-\eta\Big\{\boldsymbol{u}(\boldsymbol{R}_{i}+\boldsymbol{\rho}_{i})\,\hat{\bm{\rho}}_{i}+\hat{\bm{\rho}}_{i}\,\boldsymbol{u}(\boldsymbol{R}_{i}+\boldsymbol{\rho}_{i})\Big\}\right]d\text{S}_{i},
\end{eqnarray}
is the contribution to the stress from the particles \cite{landau1959fluid}.
From the expansion of Eq.(\ref{eq:boundaryFields-Yl}), it follow
that
\begin{eqnarray}
 & \boldsymbol{\Sigma}^{P} & =\frac{1}{V}\sum_{i}\left[b\mathbf{F}_{i}^{(2)}-\frac{8\pi\eta b^{2}}{3}\mathbf{V}_{i}^{(2s)}\right].
\end{eqnarray}
The symmetric part of the particle contribution to the bulk stress
was denoted by Batchelor as the stresslet \cite{batchelor1970stress}. 

To obtain the rheological response of the suspension, the above quantity
has to be calculated in the presence of an externally imposed flow
$\boldsymbol{u}^{{\scriptscriptstyle \infty}}$. The irreducible tensorial
harmonics coefficients of external flow
\begin{equation}
\mathbf{V^{{\scriptscriptstyle \infty}}}_{i}^{(l)}=\frac{2l-1}{4\pi b^{2}}\int\boldsymbol{u}^{{\scriptscriptstyle \infty}}(\boldsymbol{R}_{i}+\bm{\rho}_{i})\,\mathbf{Y}^{(l-1)}(\bm{\hat{\rho}}_{i})\,d\text{S}_{i},
\end{equation}
gives corresponding coefficients of the traction $\mathbf{F^{\infty}}_{i}^{(l\sigma)}$
from the solution of linear system for external flow\begin{subequations}
\begin{alignat}{1}
 & \qquad\qquad\,\,\,\,\mathbf{V}_{i}^{{\scriptscriptstyle \infty}(l)}=\boldsymbol{G}_{ij}^{(l,\,l')}\cdot\mathbf{F}_{j}^{^{\infty}(l')},\\
\mathbf{F^{\infty}}_{i}^{(l\sigma)} & =\hat{\boldsymbol{\gamma}}_{ij}^{(l\sigma,T)}\cdot\mathbf{V}_{j}^{\infty}+\hat{\boldsymbol{\gamma}}_{ij}^{(l\sigma,R)}\cdot\mathbf{\Omega}_{j}^{\infty}+\hat{\boldsymbol{\gamma}}_{ij}^{(l\sigma,l'\sigma')}\cdot\mathbf{V}_{j}^{\infty(l'\sigma')}.
\end{alignat}
\end{subequations}Here $\hat{\boldsymbol{\gamma}}^{(l\sigma,\,l'\sigma')}=\mathbf{P}^{(l\sigma)}\cdot\left[\boldsymbol{G}^{-1}\right]^{(l,\,l')}\cdot\mathbf{P}^{(l'\sigma')}$
are generalized friction tensors encoding the response to external
flow and its explicit expression is obtained by repeating the steps
in Section \ref{sec:generalized-stokes}. The particle contribution
to the suspension stress, then, is the sum of external and active
contributions: 
\begin{eqnarray}
V\mathbf{\Sigma}^{P}= & \underbrace{\sum_{l\sigma=1s}^{\infty}\left[\hat{\boldsymbol{\gamma}}_{ij}^{(2s,\,l\sigma)}\negmedspace+\tfrac{1}{2}\boldsymbol{\varepsilon}\cdot\hat{\boldsymbol{\gamma}}_{ij}^{(2a,\,l\sigma)}\negmedspace+\tfrac{\boldsymbol{I}}{3}\hat{\boldsymbol{\gamma}}_{ij}^{(2t,\,l\sigma)}\right]\cdot\mathbf{V}_{j}^{{\scriptscriptstyle \infty}(l\sigma)}}_{\mathrm{External\,flow}}-\underbrace{\sum_{l\sigma=1s}^{\infty}\left[\boldsymbol{\gamma}_{ij}^{(2s,\,l\sigma)}\negmedspace+\tfrac{1}{2}\boldsymbol{\boldsymbol{\varepsilon}}\cdot\boldsymbol{\gamma}_{ij}^{(2a,\,l\sigma)}\negmedspace+\tfrac{\boldsymbol{I}}{3}\boldsymbol{\gamma}_{ij}^{(2t,\,l\sigma)}\right]\cdot\mathbf{V}_{j}^{(l\sigma)}}_{\mathrm{Active\,slip}}.
\end{eqnarray}
All particle indices are summed over in the above. The friction tensors
in an unbounded fluid, obtained in Appendix \ref{appendix:iterative-sol},
can be used to estimate the active contribution to the suspension
stress. The above expression must be statistically averaged over the
position and orientation of the colloids to obtain the average stress
in the suspension using distribution function $\Psi(\mathbf{\boldsymbol{R}}^{N},\mathbf{\boldsymbol{p}}^{N})$
defined in Eq.(\ref{eq:dist-func}). 

We now consider a force-free, torque-free suspension and assume that
the leading contributions from the external flow produces a pure strain
on the surface of the colloids. Thus, the first symmetric traceless
moment of the external flow $\boldsymbol{v}^{{\scriptscriptstyle \infty}}$
is most-dominant and is parametrize as $\mathbf{V}_{i}^{{\scriptscriptstyle \infty}(2s)}=b\mathbf{E}$.
Expression for $\mathbf{\Sigma}^{P}$ then reduces to
\begin{align}
\mathbf{\Sigma}^{P}=\sum_{i} & \frac{20\pi\eta b^{3}}{3}\mathbf{E}-\sum_{i}\frac{28\pi\eta b^{2}}{3}\mathbf{V}_{i}^{(2s)}+O(\phi^{2}),
\end{align}
where the first two terms are the leading order one-body contribution
due to external flow and activity respectively. They are the $O(\phi)$
contribution to the suspension stress, where $\phi$ is the suspension
volume fraction. At $O(\phi)$, the average stress depends on the
average of the irreducible dipole $\langle\mathbf{V}_{i}^{(2s)}\rangle_{\{\boldsymbol{p}_{i}\}}$
over the orientational distribution function. The orientational distribution
function of spheres remains unchanged in a shear flow. In an isotropic
suspension, therefore, the average $\langle\mathbf{V}_{i}^{(2s)}\rangle_{\{\boldsymbol{p}_{i}\}}$
vanishes and there is no contribution to the suspension stress at
$O(\phi)$ due to activity, as was first pointed out by Pedley and
Ishikawa \cite{ishikawa2007rheology}. However, if the distribution
is not isotropic, then the average of $\langle\mathbf{V}_{i}^{(2s)}\rangle_{\{\boldsymbol{p}_{i}\}}$
is proportional to $V_{0}^{(2s)}$ and non-zero. In particular, if
the first symmetric moment of the orientational distribution function
is non-zero, then, the stress may increase or decrease depending on
$V_{0}^{(2s)}$ begin negative (for contractile colloids) or positive
(for extensile colloids). 

The exact relation between the suspension stress and the generalized
friction tensor obtained above can be used to derive the $O(\phi^{2})$
and higher corrections to the suspension stress. Such a calculation
requires a careful regularization of conditionally convergent integrals
\cite{o1979method} and will be presented in a future work.

\section{Active pressure in external potential\label{sec:Active-pressure}}

From the forces, torques and stresslets on the colloids discussed
in the previous sections, we now turn our attention to the mechanical
pressure in the fluid. The active contribution to the pressure in
the fluid is given by the second term in Eq.(\ref{eq:fuid-pressure-irred}).
We consider a suspension of active colloids confined by an external
spherical potential such that they are always inside a sphere of radius
$R.$ The confining surface is not a physical boundary and there is
free motion of the fluid across it. We use the results of the section
\ref{sec:Boundary-integral-solution} and \ref{sec:langevin-description}
to calculate the fluid pressure on the confining surface due to the
motion of the active colloids in the interior volume. This geometry
is motivated by the confinement of bacteria inside a fluid drop with
a porous interface.

We retain slip modes $l\sigma=2s$ and $4a$ which correspond to the
symmetric irreducible dipole and the chiral octupole. The first generates
a long-ranged flow while the second produces self-rotation \cite{ghose2014irreducible}.
For simplicity, we choose these modes to be uniaxial, parametrized
in terms of the orientation $\boldsymbol{p}_{i}$ of the colloid,
as given in Eq.(\ref{eq:uniaxial-parametrization}). The microscopic
dynamics and the pressure distributions are sensitive to the sign
of $V_{0}^{(2s)}$ as we shall see below. The fluid flow due to the
chiral term, which decays as $r_{{\scriptscriptstyle {ij}}}^{-4}$,
induces a net rotation of the system. The most dominant contribution
to the fluid flow comes from the long-ranged dipolar flow, which decays
as $r_{{\scriptscriptstyle {ij}}}^{-2}$. The fluid pressure decays
as one power higher than the fluid flow. The dipole, thus, crucially
determines the dynamic of active colloids in the spherical confinement
and fluid pressure on the confining surface.

The confining potential $U^{\mathrm{c}}(R_{i})=k^{\mathrm{c}}\exp\tfrac{1}{R-R_{i}}/(R_{{\scriptscriptstyle {max}}}-R_{i})$
for $R_{i}>R$ and $U^{\mathrm{c}}(R_{i})=0$ otherwise. Here $R_{{\scriptscriptstyle {max}}}$
is chosen to be few particle radius more than $R$ in simulations
and $k^{\mathrm{c}}$ is the strength of the potential. The colloids
also have an additional steric interaction, which is modeled by a
short ranged repulsive potential, and depends on the separation $\boldsymbol{r}_{{\scriptscriptstyle {ij}}}=\boldsymbol{\boldsymbol{R}}_{i}-\boldsymbol{\boldsymbol{R}}_{j}$
between the colloids. This potential is modeled using the WCA potential
for separation $r_{{\scriptscriptstyle {ij}}}<r_{{\scriptscriptstyle {min}}}$,
$U(r_{{\scriptscriptstyle {ij}}})=\epsilon(\frac{r_{{\scriptscriptstyle {min}}}}{r_{{\scriptscriptstyle {ij}}}})^{12}-2\epsilon(\frac{r_{{\scriptscriptstyle {min}}}}{r_{{\scriptscriptstyle {ij}}}})^{6}+\epsilon,$
and zero otherwise \cite{weeks1971role}. Here $\epsilon$ is the
strength of the potential. The specification of the slip and body
forces complete the description of our model. We start the simulations
with a completely random distribution of hard spheres positioned symmetrically
about the origin \cite{skoge2006packing}. The orientations of all
the colloids are pointing along the $\mathbf{\hat{z}}-$axis. The
tendency of the hydrodynamic torques acting on the colloids to rotate
their orientation is nullified by external torques $\mathbf{T}_{i}^{P}=T_{0}(\boldsymbol{p}_{i}\times\mathbf{\hat{z}})$
arising from bottom-heaviness. Thus the orientation of all the colloids
remain along $\mathbf{\hat{z}}-$axis for all times. We then study
the collective dynamics and measure the pressure on the ``confining''
surface. 

\begin{figure*}
\includegraphics[width=0.92\textwidth]{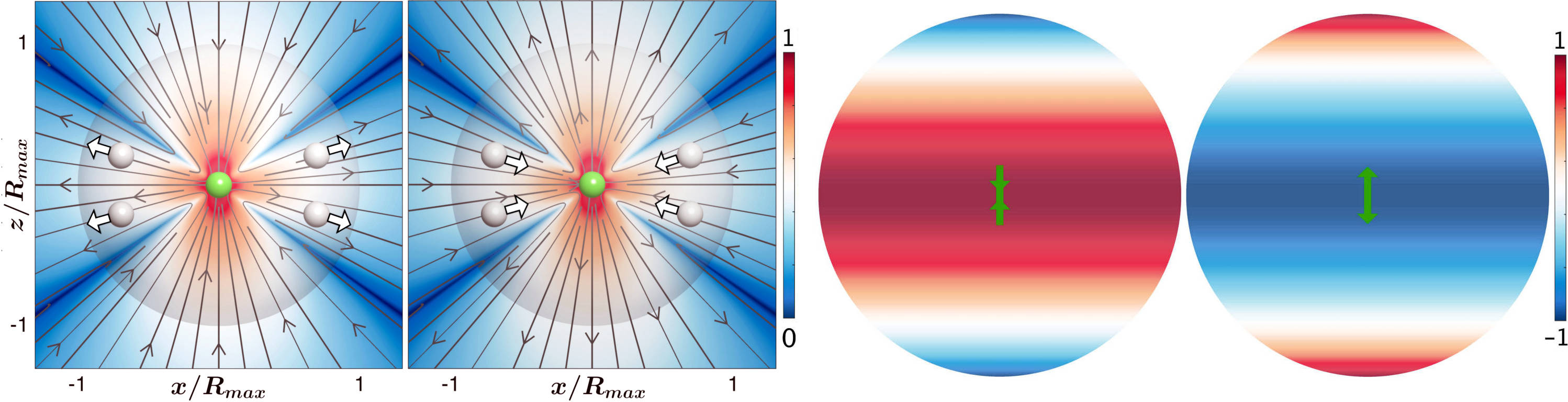}\caption{Axisymmetric flow around a contractile active colloid (left) and an
extensile active colloid (right) are shown in first two panels, while
the last two panel are the pseudo color plots of the active pressure
on a sphere enclosing the respective colloids. Streamlines of the
flow are overlaid on the pseudo color map of the normalized logarithm
of the flow speed. The source active colloid is colored in green while
colloids in white color are tracers with a solid white arrow showing
the direction of forces acting on them. \label{fig:presssure-plots}\label{fig:Flow-diagram-dipole} }
\end{figure*}

The sign of the strength of the symmetric irreducible dipole, $V_{0}^{(2s)}$,
is positive (negative) for an extensile (contractile) active colloid.
We plot the fluid flow produced by a contractile and extensile colloid
in first two panels in Fig.(\ref{fig:Flow-diagram-dipole}). The orientation
of the colloids is assumed to be along the $\hat{\mathbf{z}}$ direction.
The source colloid is colored in green, while white arrows on the
tracer colloids show the direction of the force acting on them. The
direction of the forces on these colloids gives a heuristic understanding
of the dynamics in the spherical confinement as we explain below.
The last two panels of Fig.(\ref{fig:Flow-diagram-dipole}) contain
the fluid pressure due to contractile and extensile active colloid.
It can be seen that the pressure along the equator is higher for for
contractile colloids while pressure at the poles is higher for extensile
colloids. Collective dynamics of contractile and extensile active
colloids under spherical confinement follow from the flow field of
the individual colloids.

\begin{figure*}
\includegraphics[width=0.96\textwidth]{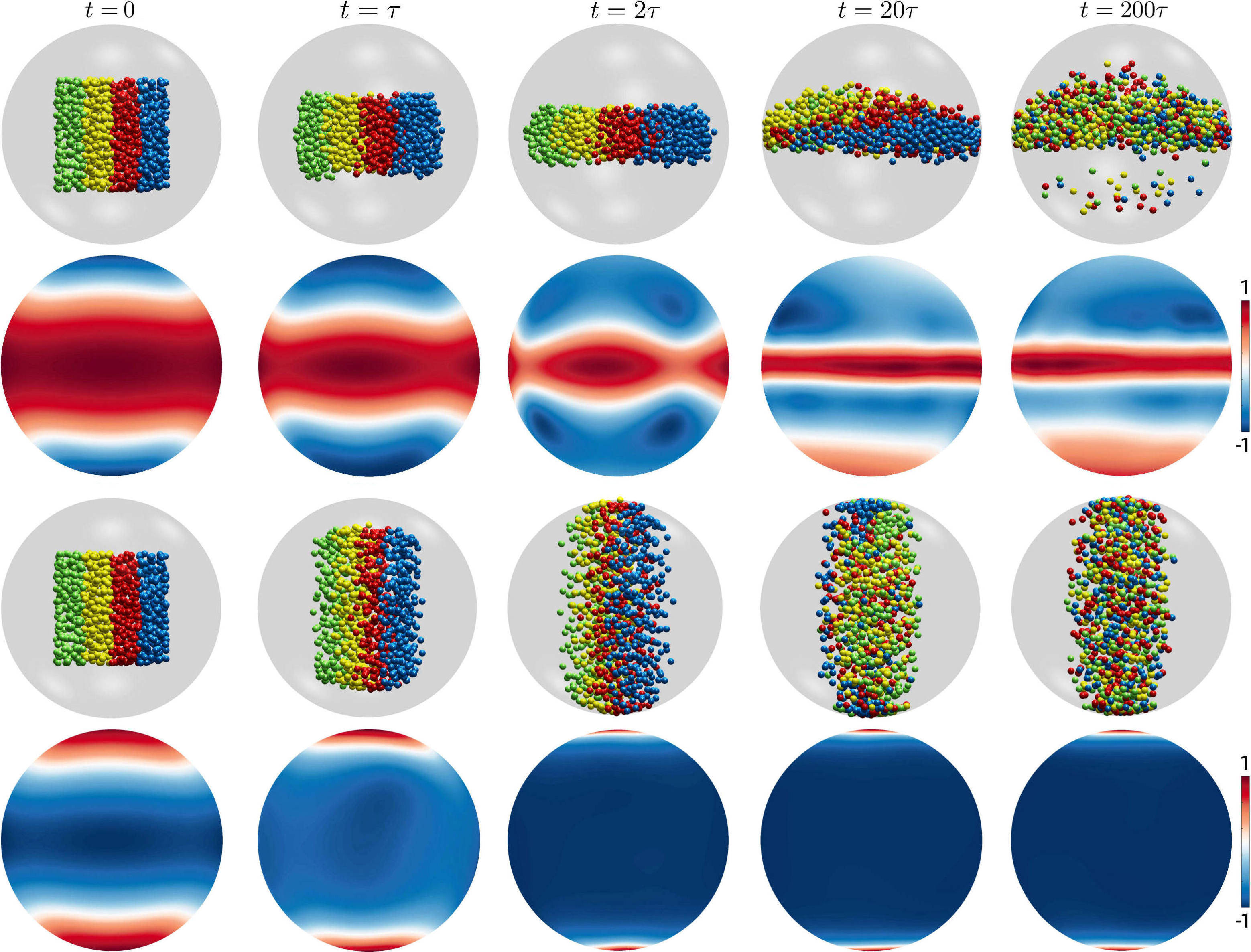}\caption{Dynamics of 1024 contractile and extensile active colloids in a spherical
confinement and the active fluid pressure on the spherical surface
confining them. The first row shows the lateral view of configurations
from a simulation of contractile active colloids, confined in a spherical
potential, at different times ($\tau=b/V_{0}^{(4a)}$). The colloids
are colored by their initial positions. The second row shows the pseudo-color
plots of the normalized active pressure on the sphere confining the
colloids at corresponding time instants. The remaining two rows are
the same set of plots but for extensile active colloids (see text).\label{fig:Dynamics-of-contractile}
\label{fig:Dynamics-of-extensile}}
\end{figure*}

A contractile colloid pushes the particles away in the plane perpendicular
to the dipole axis, which leads to an instability in an initially
isotropic suspension of colloids. The colloids final reach to a steady
state which they organize into a continuously rearranging ``oblate''
structure. The dipoles tend to push each other as far as possible
but the spherical confinement coupled with short-ranged repulsion
make them undergo rolls, which accounts for the continuous rearrangement
of the structure. In the first two rows of Fig.(\ref{fig:Dynamics-of-contractile}),
we show instantaneous configurations of contractile colloids and the
state of the active pressure on the bounding sphere. The fluid pressure
is then maximum on the equator of the confining sphere.

The last two rows of Fig.(\ref{fig:Dynamics-of-contractile}) show
the corresponding configurations and fluid pressure on a confining
sphere for extensile dipoles. The initial isotropic distribution of
the colloids finally finds a steady state in a ``prolate'' distribution
of of extensile dipoles. The colloids continuously rearrange this
structure as the dipolar flow tries to push them apart while the confining
sphere holds them back. Moreover, the pressure is higher at the poles
in this case. 

To summarize, we have shown that the dynamics and fluid pressure measured
in a suspension of extensile and contractile active colloids are completely
\textit{different.} Movie 1 and 2 of \cite{supplementalTraction}
show the dynamics of contractile and extensile colloids, respectively,
in a spherical confinement. The reported dynamics continues to hold
for a random initial distribution with arbitrary separation between
the colloids, and even at a very small initial volume fraction. This
is because of the long-range $(1/r^{2})$ attractive forces, between
the colloids, driving their aggregation. Here, we have assumed that
the orientations of colloids are fixed by their bottom-heaviness.
A more detailed study of the given system, where this condition is
relaxed, will be pursued in a future work.

\section{Dynamics in an optical lattice\label{sec:Dynamics-optical}}

The simplest system in which an interplay of non-uniform external
fields, activity and Brownian motion can be studied is an active colloid
confined in a three-dimensional harmonic potential. As this system
is both experimentally realizable in optical trapping experiments
\cite{jiang2010active,moyses2016trochoidal} and analytically tractable
\cite{tailleur2008statistical,nash2010,singh2015many} it serves as
the ``Ising model'' of active colloidal physics. In this section,
we use the Langevin equations to study the motion of active colloids
confined in a square array of harmonic confining potentials. The principal
question we focus on is the collective dynamics of the colloids and,
in particular, their synchronization. The system we study can be realized
experimentally in holographic tweezers. From estimates of the dimensionless
groups presented in the previous section, it is clear that Brownian
motion can be ignored in the first approximation. Accordingly, we
neglect noise and study the mechanics of the system, postponing the
study of its statistical mechanics to a future work.

We consider self-propelling, polar, achiral active colloids, with
non-zero values of $\mathbf{V}_{i}^{\mathcal{A}}$, \textbf{$\mathbf{V}_{i}^{(2s)}$}
and\textbf{ $\mathbf{V}_{i}^{(3t)}$}. With this choice, an isolated
colloid translates with velocity $\mathbf{V}_{i}^{\mathcal{A}}=v_{s}\boldsymbol{p}_{i}$
while producing dipolar and quadrupolar flows of strengths proportional
to $\mathbf{V}_{i}^{(2s)}$ and \textbf{$\mathbf{V}_{i}^{(3t)}$}
respectively \cite{ghose2014irreducible,singh2015many}. The centers
of the $N$ traps are at $\boldsymbol{R}_{i}^{0}$, arranged linearly
or in a $\sqrt{N}\times\sqrt{N}$ square lattice. Each trap contains
a single active colloid which feels a body force from that trap alone.
The moment of force about the trap center is zero. Therefore, in a
trap of stiffness $k$ centered at \textbf{$\boldsymbol{R}_{i}^{0}$}
\begin{equation}
\mathbf{F}_{i}^{P}=-k(\boldsymbol{R}_{i}-\boldsymbol{R}_{i}^{0}),\qquad\mathbf{T}_{i}^{P}=0.
\end{equation}
First, we consider the dynamics ignoring hydrodynamic interactions.
Then, force and torque balance give
\begin{eqnarray}
-6\pi\eta b\cdot(\mathbf{V}_{i}-v_{s}\boldsymbol{p}_{i})-k(\boldsymbol{R}_{i}-\boldsymbol{R}_{i}^{0}) & =0 & ,\qquad-8\pi\eta b^{3}\,\mathbf{\Omega}_{i}=0.
\end{eqnarray}
In the absence of hydrodynamic and Brownian torques, there is no angular
velocity and the colloid translates in a direction $\boldsymbol{p}_{i}$
chosen by the initial condition. It is brought to rest at a radius
$R^{*}=6\pi\eta bv_{s}/k=\mathcal{A}_{T}b$, when the propulsive and
trap forces are balanced \cite{nash2010,singh2015many}. Here $\mathcal{A}_{T}$
is the activity number defined in Eq.(\ref{eq:dimls-number}), which
quantifies the ratio of active propulsive force to the passive confining
force. The stationary state, without hydrodynamic interactions, is
one in which all colloids are confined at a distance $R^{*}$ from
the center of its trap and oriented radially outward in a direction
that is, in general, different for each colloid.
\begin{figure*}
\includegraphics[width=1\textwidth]{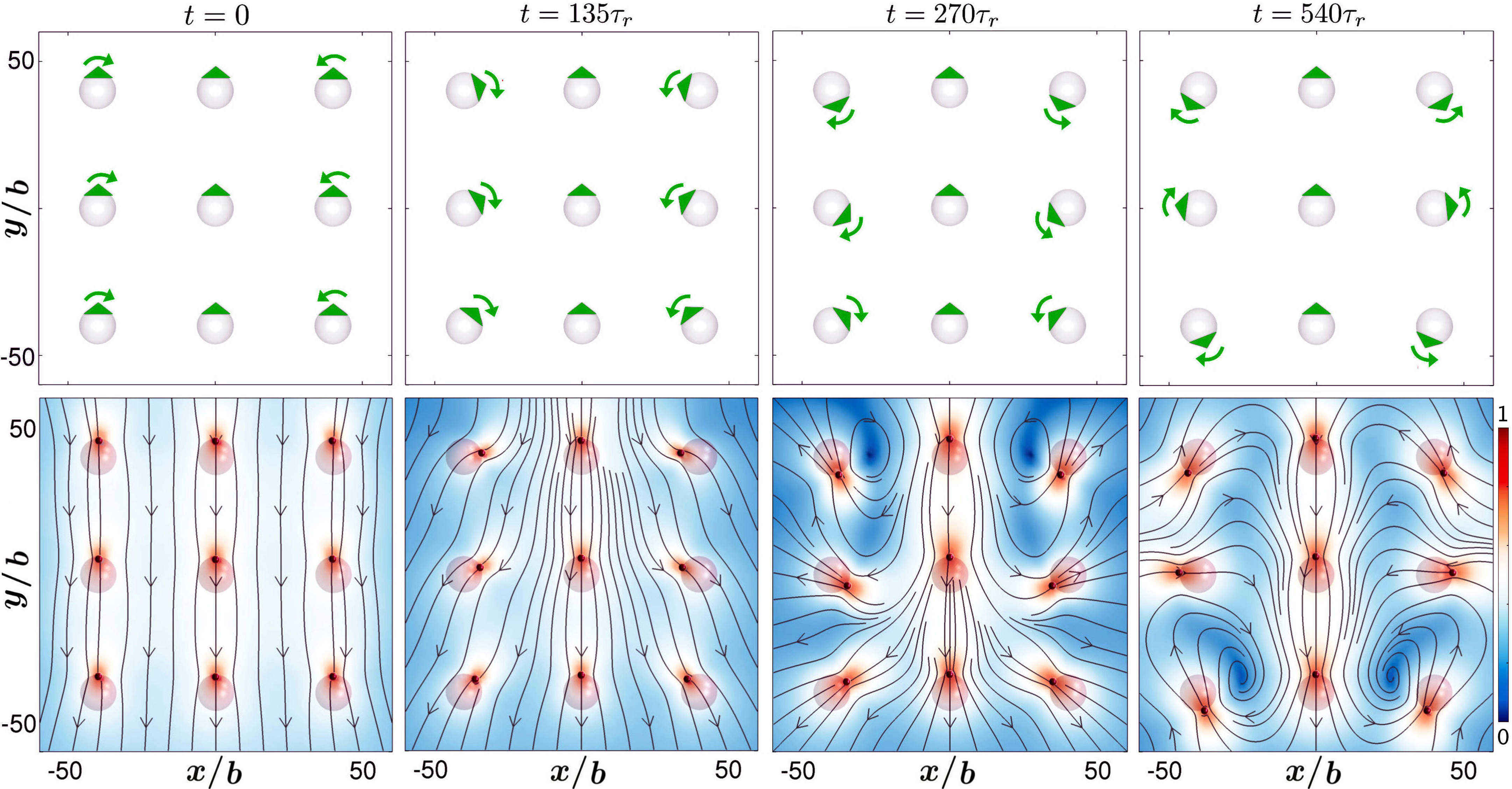}\caption{Instantaneous configurations from simulations of active colloids in
$3\times3$ square lattice of harmonic traps (top panel) and corresponding
streamlines of fluid flow (bottom-panel) overlaid on the pseudo-color
plot of the normalized logarithmic flow speed at different times ($\tau_{r}=8\pi\eta b/k$).
The traps are shown by the schematics spheres while the positions
and orientations of the colloids are shown by green cones, and curved
green arrows show rotation. The colloids on the symmetry axis do not
rotate, while those on the left and right rotate clockwise and counter-clockwise,
respectively, in a synchronized manner.\label{fig:3x3-traps}}
\end{figure*}
\begin{figure*}[t]
\includegraphics[width=0.98\textwidth]{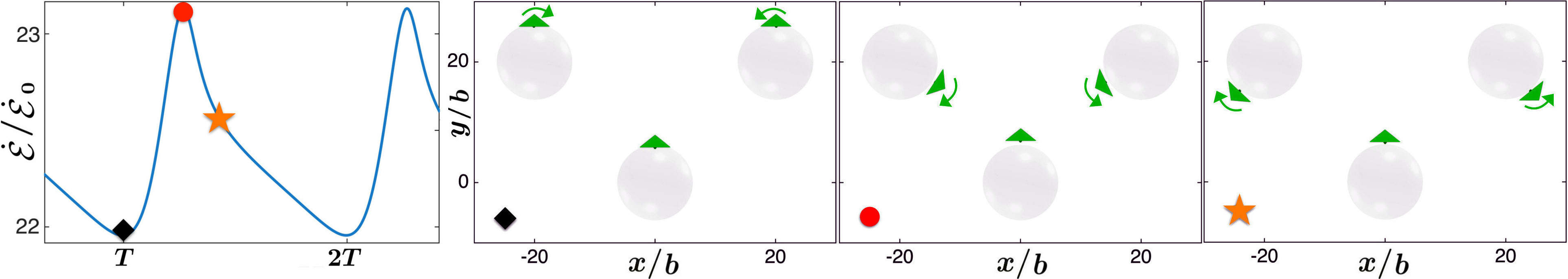}\caption{Power dissipation for the synchronized and periodic motion of three
active colloids in a triangular lattice of harmonic traps (left panel),
scaled by power dissipation of an isolated colloid $\dot{\mathcal{E}}_{0}$.
The remaining plots show configuration of the system at three instants
(shown by corresponding markers) along the power dissipation curve.
The traps are shown by the schematics spheres and the position and
orientation of the colloids are shown by green cones, and curved green
arrows show rotation.\label{fig:power-dissipation-three-traps}}
\end{figure*}

This state is destabilized \emph{with} hydrodynamic interactions \cite{nash2010,hennes2014self,singh2015many}
due to the torque induced by the flow of the $l\sigma=1s$ and $2s$
modes \cite{singh2015many}. The leading contributions to the hydrodynamic
torque is
\begin{equation}
\mathbf{T}_{i}^{H}=-\boldsymbol{\gamma}_{ij}^{RT}\cdot(\mathbf{V}_{i}-v_{s}\boldsymbol{p}_{i})-\boldsymbol{\gamma}_{ij}^{(R,\,2s)}\cdot\mathbf{V}_{j}^{(2s)},
\end{equation}
which, upon using the explicit forms of the generalized friction tensors
from Appendix \ref{appendix:iterative-sol}, is
\begin{align}
\mathbf{T}_{i}^{H} & =8\pi\eta b^{3}\bigg[-\frac{\hat{\boldsymbol{r}}_{{\scriptscriptstyle ij}}}{r_{{\scriptscriptstyle {ij}}}^{2}}\times\frac{k(\boldsymbol{R}{}_{j}-\boldsymbol{R}{}_{j}^{0})}{6\pi\eta}+\frac{14(\boldsymbol{p}_{j}\cdot\hat{\boldsymbol{r}}_{{\scriptscriptstyle ij}})\,}{r_{{\scriptscriptstyle {ij}}}^{3}}(\boldsymbol{p}_{j}\times\hat{\boldsymbol{r}}_{{\scriptscriptstyle ij}})V_{0}^{(2s)}b^{2}\,\bigg]+O(r_{{\scriptscriptstyle {ij}}}^{-3}).
\end{align}
Here $\boldsymbol{r}_{{\scriptscriptstyle {ij}}}=\boldsymbol{\boldsymbol{R}}_{i}-\boldsymbol{\boldsymbol{R}}_{j}$
and the force balance equation has been used to eliminate $\mathbf{V}_{i}-v_{s}\boldsymbol{p}_{i}$
in favor of the trapping force. The hydrodynamic torque vanishes when
the colloids are collinear and their orientations are along the line
joining their centers. Thus stable states of rest are possible even
in the presence of hydrodynamic interactions for specially chosen
initial conditions \cite{singh2015many}. In general, though, the
interplay of self-propulsion, confinement, and hydrodynamic interactions
produce steady states with continuous motion. 

With this understanding, we now present numerical results for dynamics
in a lattice of traps. In a linear lattice of traps, we find stable
stationary states, reached irrespective of initial conditions, in
which all colloids are oriented \textit{along} the line joining the
trap centers and at a confinement radius that is slightly altered
from $R^{*}$ due to hydrodynamic interactions. We then study dynamics
in a $3\times3$ square lattice of traps. The dynamics is shown in
Fig.(\ref{fig:3x3-traps}) and Movie 3 of \cite{supplementalTraction}.
The initial condition is chosen to be a stable state in the absence
of hydrodynamic interactions. We find that the particles at the center
do not rotate by symmetry while particles on the left of this symmetry
axis, rotate clockwise and particle on their right rotate counter-clockwise.
This can be understood by the estimating the hydrodynamic torques
on each colloid. The dynamics can also understood intuitively from
the flow field of Fig.(\ref{fig:3x3-traps}). The colloids at an equal
distance from the symmetric plane have \textit{synchronized} dynamics.
This leads to a long-ranged correlation between the colloids. In summary,
(a) there is a rotational instability in the system if the traps centers
are not collinear, and (b) dynamics is synchronized about an axis
of symmetry in non-collinear traps. 

In Fig.(\ref{fig:power-dissipation-three-traps}), we estimate the
power dissipation for the synchronized and periodic motion of three
active colloids in a triangular lattice of harmonic traps. The dynamics
in a triangular lattice of harmonic traps is similar to that of the
square lattice of traps. The colloid in the central trap has no rotational
dynamics due to symmetry, while the one on the left rotates clockwise
and the colloid on the right rotates counter-clockwise. The power
dissipated in the system is minimum when the colloids are widely separated,
while the dissipated power is maximum when they are closer to each
other. Thus, the first configuration of Fig.(\ref{fig:power-dissipation-three-traps}),
corresponds to the minimum of power dissipation, while the second
is the maximum, as indicated by the markers on the power dissipation
curve, and the third configuration is an intermediate value.

The motion of many hydrodynamically interacting active colloids in
a harmonic potential was first studied by Nash \emph{et al }\cite{nash2010}
using lattice Boltzmann simulations and then by singularity \cite{hennes2014self}
and boundary integral \cite{singh2015many} methods. There, the interplay
between self-propulsion, confinement and hydrodynamically-induced
reorientation yields orbits for a pair of confined particles \cite{singh2015many}.
These individual orbits coalesce to produce sustained convection in
a confined suspension \cite{singh2015many} producing the so-called
`self-assembled pump' \cite{nash2010}. In contrast, here we study
the motion of many active colloids in a lattice of harmonic traps
to uncover strikingly different dynamics. The numerical simulations
of this work are performed using the PyStokes library \cite{pystokes}.

\section{Discussion and summary\label{sec:Discussion-and-summary}}

By exploiting the linearity of slow viscous flow, as manifest in its
boundary integral representation, we have derived linear relations
between the coefficients of the force per unit area and the active
slip in a suspension of active colloidal spheres. These linear relations
we call the ``generalized Stokes laws'' and the tensorial coefficients
relating the force per unit area to the slip we call the generalized
friction tensors. We have derived explicit expressions for these tensors
in terms of the Green's function of Stokes flow. The boundary integral
representation provides the flow at any point in the bulk fluid, given
the force per unit area and the active slip on the fluid-colloid boundary.
This leads to numerical methods that are more efficient than those
that need to resolve the bulk fluid flow. \cite{nash2010,jayaraman2012,delmotte2015large}.

From the generalized Stokes laws, we directly obtain the forces, torques,
fluid flow, fluid pressure, power dissipation and suspension stress.
Since forces and torques are the fundamental dynamical quantities
in Newtonian or Langevin descriptions of particle dynamics, our contribution
forms the basis for a microscopic theory of active suspension mechanics
and statistical mechanics that conserves momentum in both the bulk
fluid and at fluid-solid boundaries. The formalism is applied to experimentally
realizable situations to derive testable predictions. It should be
noted that in this work, the simulations are done in the mobility
formulation and the leading terms of the forces and torques are calculated
using the quick and transparent Jacobi iterative scheme. When greater
accuracy is desired, conjugate gradients or other Krylov subspace
methods may be used \cite{hestenes1952methods}. 

The Langevin and Smoluchowski description of active colloids are obtained
in terms of mobility matrices and the propulsion tensors. The far-field
limit of the mobility matrices are obtained in terms of the Green's
function of the Stokes flow while a lubrication approximation may
be used when the colloids are close to each other \cite{brady1988dynamic}.
The propulsion tensors are obtained in terms of the lubrication-corrected
mobilities and the friction tensors. Thus we account for both the
far-field hydrodynamic interactions, to any order of desired accuracy,
and the near-field lubrication interactions. The Galerkin discretization
of the boundary integral equation provides most accurate results for
smooth boundaries, like spheres, for least number of unknowns and
preserves the self-adjointness of the problem \cite{muldowney1995spectral,youngren1975stokes}.
Dynamic simulation of hundreds of thousand of active colloids on a
multi-core computational architectures is possible. 

In this work, we assume a spherical particle with active slip on its
surface, which is then expanded in a Galerkin basis to obtain the
force per unit area. Thus, any generic mechanism generating the active
velocity can be modeled in our approach. Typically, the slip mechanism
for synthetic active colloids is phoretic and then, we do need to
solve separately for a concentration field \cite{uspal2015self}.
Here, we have assumed that the non-hydrodynamic parts of the problem
have been solved separately. This assumption requires the decoupling
of advection and diffusion and, therefore, is restricted to low Péclet
numbers. 

It is useful to compare the results presented here with existing results
for hydrodynamic interactions of many spheres. Excluding the contribution
from active slip and truncating the Galerkin expansion at $l=2$,
results in the method of computing far-field hydrodynamic interactions
in the so-called FTS Stokesian dynamics method of Brady and coworkers
\cite{durlofsky1987dynamic,brady1988dynamic}. This latter method
ignores the entire $l=3$ contribution which decays as $r^{-3}$ for
unbounded flow and is long-ranged. This low-order truncation has been
extended up to to $7$-th order by Ichiki \cite{ichiki2002improvement}.
The method uses different bases for the expansion of the force per
unit area and the surface velocity and several elaborate transformations
are necessary to obtain a full-rank linear system. In contrast, our
choice of identical basis for both force per unit area and active
slip automatically yields a full-rank linear system. In our basis,
harmonics indexed by $l$ produce bulk flows that decay as $r^{-l}$
, a simplicity that is absent in the bases used by Cichocki \cite{cichocki1994friction}
and Ichiki \cite{ichiki2002improvement}. For active colloids, earlier
work closest in spirit to ours is that of Ishikawa \emph{et al }\cite{ishikawa2006,ishikawa2008development,kyoya2015shape}
where axisymmetric slip velocities, truncated to the first two non-trivial
modes, are considered. The far-field and near-field hydrodynamic interactions
are obtained, respectively, in superposition and lubrication approximations.
In contrast, we include the most general form of the slip and use
an irreducible basis function for Galerkin discretization which gives
a systematic way of evaluating hydrodynamic interactions to any desired
order of accuracy. Active colloids can be studied by other models
like the force-coupling method, though only in dilute limit, as the
distinction between the interior and exterior of a colloid is notional
in the method \cite{delmotte2015large}. We refer the readers to recent
reviews on low Reynolds number flows for a more comprehensive list
of work in the field \cite{zottl2016emergent,goldstein2016batchelor,lauga2009,winkler2016low}.

The present paper focuses on the deterministic parts of the problem,
though the stochastic equations have been presented. In a future work,
we will explore the stochastic aspects more fully, using the Langevin
and Smoluchowski description derived in this paper. The formalism
can also be extended to study fluctuations in chains of active particles
\cite{jayaraman2012,laskar2013,laskar2015brownian,pandey2016flow}.
Applications of our method to collective phenomena in magnetotactic
colloids and to active rheology will be presented in forthcoming work
as will be the extension to ellipsoidal particles.
\begin{acknowledgments}
We thank M. E. Cates, S. Ghose, A. Laskar, R. Manna, H. A. Stone and
G. Subramanian for many useful discussions. Financial support from
the Department of Atomic Energy, Government of India, and computing
resources through Annapurna and Nandadevi cluster at the Institute
of Mathematical Sciences is gratefully acknowledged. 
\end{acknowledgments}

\appendix

\section{Iterative solution for generalized friction tensors\label{appendix:boundary-integrals}\label{appendix:iterative-sol}}

In this section, we derive iterative solutions of the generalized
friction tensors using the properties of the tensorial spherical harmonics
$\mathbf{Y}^{(l)}(\bm{\hat{\rho}}_{i})$. The first four $\mathbf{Y}^{(l)}s$
are
\begin{equation}
Y^{(0)}=1,\qquad Y_{\alpha}^{(1)}=\hat{\rho}_{\alpha},\qquad Y_{\alpha\beta}^{(2)}=\left(\hat{\rho}_{\alpha}\hat{\rho}_{\beta}-\tfrac{\delta_{\alpha\beta}}{3}\right),\qquad Y_{\alpha\beta\gamma}^{(3)}=\left(\hat{\rho}_{\alpha}\hat{\rho}_{\beta}\hat{\rho}_{\gamma}-\tfrac{1}{5}[\hat{\rho}_{\alpha}\delta_{\beta\gamma}+\hat{\rho}_{\beta}\delta_{\alpha\gamma}+\hat{\rho}_{\gamma}\delta_{\alpha\beta}]\right).
\end{equation}
Tensorial spherical harmonics are orthogonal basis function on the
surface of a sphere
\begin{equation}
\frac{1}{4\pi b^{2}}\int\mathbf{Y}^{(l)}(\widehat{\bm{\rho}})\,\mathbf{Y}^{(l')}(\widehat{\bm{\rho}})d\text{S}=\delta_{ll'}\,\frac{l!\,(2l-1)!!}{(2l+1)}\mathbf{\Delta}^{(l)}.\label{eq:orthogonality}
\end{equation}

The expansion of the velocity in this basis has been given in the
main text. The orthogonality of the basis functions can be used to
obtain the expansion coefficients in terms of surface integrals of
traction and velocity as \cite{ladd1988,ghose2014irreducible} 
\begin{alignat}{1}
\mathbf{F}_{i}^{(l)} & =\frac{1}{(l-1)!(2l-3)!!}\int\boldsymbol{f}(\boldsymbol{R}_{i}+\bm{\rho}_{i})\mathbf{Y}^{(l-1)}(\bm{\hat{\rho}}_{i})d\text{S}_{i},\qquad\mathbf{V}_{i}^{(l)}=\frac{2l-1}{4\pi b^{2}}\int\boldsymbol{v}^{\mathcal{A}}(\boldsymbol{R}_{i}+\bm{\rho}_{i})\mathbf{Y}^{(l-1)}(\bm{\hat{\rho}}_{i})d\text{S}_{i}.
\end{alignat}
 The coefficients of the traction and velocity are tensors of rank
$l$ and can be written as irreducible tensor of rank $l,$ $l-1$
and $l-2$ \cite{singh2015many}. The decomposition of traction and
slip coefficients is given as \cite{brunn1976effect,singh2015many,schmitz1980force}\begin{subequations}
\begin{eqnarray}
\mathbf{F}_{i}^{(l)} & = & \mathbf{F}_{i}^{(ls)}-\frac{l-1}{l}\boldsymbol{\Delta}^{(l-1)}\cdot\big(\boldsymbol{\varepsilon}\cdot\mathbf{F}_{i}^{(la)}\big)+\frac{l(l-1)}{2(2l-1)}\boldsymbol{\Delta}^{(l-1)}\cdot\big(\boldsymbol{\delta}\mathbf{F}_{i}^{(lt)}\big),\\
\mathbf{V}_{i}^{(l)} & = & \mathbf{V}_{i}^{(ls)}-\frac{l-1}{l}\boldsymbol{\Delta}^{(l-1)}\cdot\big(\mathbf{\boldsymbol{\boldsymbol{\varepsilon}}}\cdot\mathbf{V}_{i}^{(la)}\big)+\frac{l(l-1)}{2(2l-1)}\boldsymbol{\Delta}^{(l-1)}\cdot\big(\boldsymbol{\delta}\mathbf{V}_{i}^{(lt)}\big).
\end{eqnarray}
\end{subequations}

The iterative solutions for generalized friction tensors are obtained
in terms of the matrix elements of the linear system Eq.(\ref{eq:linear-system}).
The key idea to obtain the exact solution of the matrix elements is
to Taylor expand the Green's function about the center of the sphere,
express the \textit{l}-th degree polynomial of the radius vector in
terms of $\mathbf{Y}^{(l)}s$, and use their orthogonality and biharmonicity
of the Green\textquoteright s function \cite{singh2015many}. Explicitly,
the matrix elements are \cite{singh2015many}
\begin{alignat}{1}
\boldsymbol{G}_{ij}^{(l,\,l')}(\boldsymbol{R}_{i},\boldsymbol{R}_{j}) & =\begin{cases}
{\displaystyle \frac{(2l-1)(2l'-1)}{(4\pi b^{2})^{2}}\int\mathbf{Y}^{(l-1)}(\hat{\bm{\rho}}_{i})\mathbf{G}(\boldsymbol{R}_{i}+\bm{\rho}_{i},\boldsymbol{R}_{j}+\bm{\rho}_{j})\mathbf{Y}^{(l'-1)}(\hat{\bm{\rho}}_{j})\,d\text{S}_{i}d\text{S}_{j};} & \qquad\quad\:{\displaystyle j=i,}\\
\, & \,\\
{\displaystyle b^{l+l'-2}\mathcal{F}_{i}^{l-1}\mathcal{F}_{j}^{l'-1}\bm{\nabla}_{{\scriptscriptstyle \boldsymbol{R}_{i}}}^{(l-1)}\bm{\nabla}_{{\scriptscriptstyle \boldsymbol{R}_{j}}}^{(l'-1)}\mathbf{G}(\boldsymbol{R}_{i},\boldsymbol{R}_{j});} & {\displaystyle \qquad\quad\;j\neq i,}
\end{cases}\\
\boldsymbol{K}_{ij}^{(l,\,l')}(\boldsymbol{R}_{i},\boldsymbol{R}_{j}) & =\begin{cases}
\frac{(2l-1)}{4\pi b^{2}(l'-1)!(2l'-3)!!}\int\mathbf{Y}^{(l-1)}(\hat{\bm{\rho}}_{i})\mathbf{K}(\boldsymbol{R}_{i}+\bm{\rho}_{i},\boldsymbol{R}_{j}+\bm{\rho}_{j})\cdot\bm{\hat{\rho}}_{j}\mathbf{Y}^{(l'-1)}(\hat{\bm{\rho}}_{j})\,d\text{S}_{i}d\text{S}_{j};\qquad & {\displaystyle j=i,}\\
\, & \,\\
{\displaystyle \frac{4\pi b^{(l+l'-1)}}{(l'-2)!(2l'-1)!!}\mathcal{F}_{i}^{l-1}\mathcal{F}_{j}^{l'-1}\bm{\nabla}_{{\scriptscriptstyle \boldsymbol{R}_{i}}}^{(l-1)}\bm{\nabla}_{{\scriptscriptstyle \boldsymbol{R}_{j}}}^{(l'-2)}\mathbf{K}(\boldsymbol{R}_{i},\boldsymbol{R}_{j});}\qquad\qquad\qquad\qquad\qquad\qquad & {\displaystyle j\neq i,}
\end{cases}
\end{alignat}
where $\mathcal{F}_{i}^{l}=\left(1+\frac{b^{2}}{4l+6}\nabla_{{\scriptscriptstyle \boldsymbol{R}_{i}}}^{2}\right)$
is an operator which encodes the finite size of the sphere \cite{ghose2014irreducible,singh2015many}. 

We now write the linear system of equations, Eq.(\ref{eq:linear-system}),
in terms of the irreducible modes of velocity and traction coefficients
as we explain now. We use the projection operator $\mathbf{P}^{(l\sigma)}$,
defined in Eq.(\ref{eq:project-vlsigma}), which projects the $\sigma$th
component of the traction and velocity coefficients, to evaluate the
generalized friction tensors $\boldsymbol{\gamma}_{ij}^{(l\sigma,\,l'\sigma')}$.
The linear system of equations is then obtained in terms of these
irreducible traction and velocity coefficients as,
\begin{alignat}{1}
\tfrac{1}{2}\mathbf{V}_{i}^{(l\sigma)}= & -\boldsymbol{G}_{ij}^{(l\sigma,\,l'\sigma')}(\boldsymbol{R}_{i},\boldsymbol{R}_{j})\cdot\mathbf{F}_{j}^{(l'\sigma')}+\boldsymbol{K}_{ij}^{(l\sigma,\,l'\sigma')}(\boldsymbol{R}_{i},\boldsymbol{R}_{j})\cdot\mathbf{V}_{j}^{(l'\sigma')}.\label{eq:linear-system-2}
\end{alignat}
Here the first two modes of $\mathbf{V}^{(l\sigma)}$ include rigid
body motion such that $\mathbf{V}^{(1s)}=\mathbf{V}-\mathbf{V}^{\mathcal{A}}$
and $\mathbf{V}^{(2a)}=b(\boldsymbol{\Omega}-\boldsymbol{\Omega}^{\mathcal{A}})$.
Having obtained the matrix elements and the linear system in the irreducible
form, we, now, provide a constructive solution for the generalized
friction tensor, given the matrix elements of the single-layer and
double-layer integrals. The solution employs the classical Jacobi
iteration, which is the mathematical basis of the physically motivated
``method of reflections'' first used by Smoluchowski \cite{ichiki2001many}.
Other methods with better convergence properties are an avenue for
further study. The generalized friction tensor is the solution to
the linear system in Eq.(\ref{eq:linear-system}), which, written
in terms of the irreducible tensorial coefficients and arranged into
standard form, is
\begin{equation}
\boldsymbol{\gamma}_{ij}^{(l\sigma,\,l'\sigma')}=\left[\boldsymbol{G}^{-1}\right]_{ik}^{(l\sigma,\,l''\sigma'')}\cdot\left(\tfrac{1}{2}\boldsymbol{I}_{kj}^{(l''\sigma'',\thinspace l'\sigma')}-\boldsymbol{K}_{kj}^{(l''\sigma'',\,l'\sigma')}\right)=\left[\boldsymbol{G}^{-1}\right]_{ik}^{(l\sigma,\,l'\sigma')}\cdot\left(\boldsymbol{B}_{kj}^{(l''\sigma'',\,l'\sigma')}\right).
\end{equation}
The right hand side of the linear system consists of the known slip
coefficients. Jacobi's solution of the friction tensor at the $n$th
iteration, for this problem, is then
\begin{alignat}{1}
\Big(\boldsymbol{\mathbf{\gamma}}_{ij}^{(l\sigma,\,l'\sigma')}\Big)^{[n]} & =\frac{1}{G_{ii}^{(l\sigma,\,l\sigma)}}\bigg[\boldsymbol{B}_{ij}^{(l\sigma,\,l'\sigma')}-\boldsymbol{\sum^{\prime}}\boldsymbol{G}_{ik}^{(l\sigma,\,l''\sigma'')}\cdot\Big(\boldsymbol{\gamma}_{kj}^{(l''\sigma'',\,l'\sigma')}\Big)^{[n-1]}\bigg].\label{eq:JacobiIteration}
\end{alignat}
The second term above is proportional to the solution obtained at
the $(n-1)$-th iteration. Here the prime over the summation indicates
that the diagonal term, corresponding to ($i=j=k$ and $l\sigma=l'\sigma'$)
is not included in the summation as per the definition of the Jacobi
iteration \cite{saad2003iterative}. The iteration must begin with
an initial guess for the solution. As the linear system is diagonally
dominant, the one-body solution is always a good starting guess. Thus,
explicit expressions for the generalized friction tensors are obtained
in terms of a Green's function of the Stokes flow. Thus the solution
is applicable to arbitrary geometries of Stokes flow. A list of Green\textquoteright s
functions of Stokes equation is given in Table (\ref{tab:geometries-G}). 

\section{Generalized friction tensors in an unbounded domain}

The one-body solution of the linear system can be calculated \emph{exactly}
in an unbounded domain. The generalized friction matrix is fully diagonal
and the Jacobi iteration trivially converges. The solution of the
linear system is then
\begin{equation}
\mathbf{F}_{i}^{(l\sigma)}=-\gamma_{ii}^{(l\sigma,\,l\sigma)}\,\mathbf{V}_{i}^{(l\sigma)}=-\frac{1}{G_{ii}^{(l\sigma,\,l\sigma)}}\mathbf{V}_{i}^{(l\sigma)}.
\end{equation}
 The first three terms of this \textit{exact} solution for one-body
system and zeroth order guess for a many-body system are
\begin{align}
\mathbf{F}_{i}^{H}=-6\pi\eta b\left(\mathbf{V}_{i}-\mathbf{V}_{i}^{\mathcal{A}}\right),\,\,\,\,\mathbf{T}_{i}^{H}=-8\pi\eta b^{3}(\mathbf{\Omega}_{i}-\mathbf{\Omega}_{i}^{\mathcal{A}}),\,\,\,\,\mathbf{F}_{i}^{(2s)}=-\frac{20\pi\eta b}{3}\mathbf{V}_{i}^{(2s)}.
\end{align}
implying that, under force-free, torque-free conditions colloids translate
and rotate independently with linear and angular velocities determined
by Eq.(\ref{eq:one-body}). This exact result for one-body motion
was obtained previously by several authors from a direct solution
of the Stokes equation \cite{lighthill1952,blake1971a,anderson1989colloid},
through the use of the reciprocal identity \cite{stone1996} and from
the boundary integral approach \cite{ghose2014irreducible,singh2015many}. 

\begin{figure*}[t]
\includegraphics[width=0.96\textwidth]{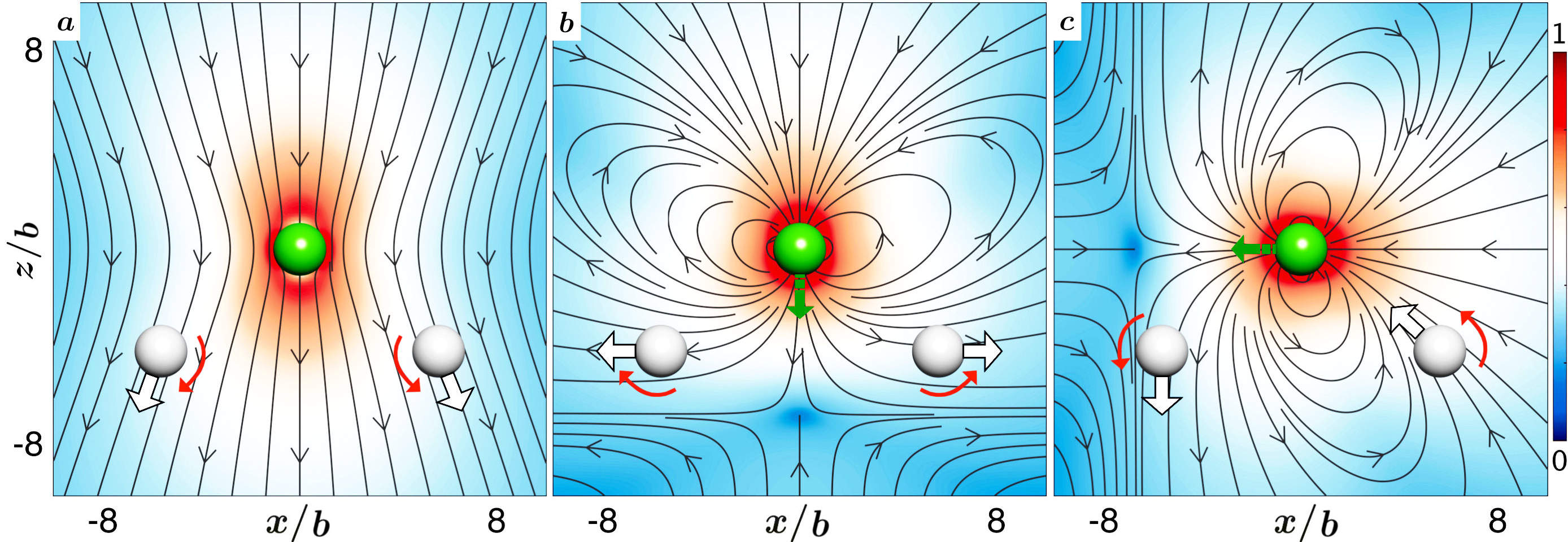}\caption{Forces and torques depend on orientation as friction is both non-local
and orientation-dependent. We demonstrate this from the streamlines
of flow around a colloid (green), while the two tracers (white) show
the direction of forces (white arrows) and torques (curved red arrows).
Panel (a) has the flow due to an external force on the colloid in
negative $\mathbf{\hat{z}}$ direction, while panel (b) and (c) contain
the flow due to active slip modes $l\sigma=2s$ and $l\sigma=3t$.
The forces and torques on the tracer colloids due to the slip modes,
panel (b) and (c), depends on the orientation (green arrow) of the
source colloid.\label{fig:orientation-dependence}}
\end{figure*}

Hydrodynamic interactions appear at \emph{first} iteration, represented
by the off-diagonal ($i\neq j$) terms in the generalized friction
tensor. The first order approximation to friction tensors in term
of a Green's function $\mathbf{G}$ of Stokes flow is
\begin{eqnarray*}
\Big(\boldsymbol{\gamma}_{ij}^{TT}\Big)^{[1]}\quad\,= & -\gamma^{T}\gamma^{T}\,\mathcal{F}_{i}^{0}\mathcal{F}_{j}^{0}\,\mathbf{G},\qquad\quad\qquad\qquad\quad\qquad\qquad & \Big(\boldsymbol{\gamma}_{ij}^{RT}\Big)^{[1]}\quad\,=\,-\frac{1}{2}\,\gamma^{T}\gamma^{R}\,\boldsymbol{\nabla}_{{\scriptscriptstyle \boldsymbol{R}_{i}}}\times\mathbf{G},\\
\Big(\boldsymbol{\gamma}_{ij}^{TR}\Big)^{[1]}\quad\,= & -\cfrac{1}{2}\,\gamma^{T}\gamma^{R}\,\boldsymbol{\nabla}_{{\scriptscriptstyle \boldsymbol{R}_{j}}}\times\mathbf{G},\quad\quad\,\,\,\,\qquad\qquad\qquad\qquad & \Big(\boldsymbol{\gamma}_{ij}^{RR}\Big)^{[1]}\quad\,=\,-\frac{1}{4}\,\gamma^{R}\gamma^{R}\,\boldsymbol{\nabla}_{{\scriptscriptstyle \boldsymbol{R}_{i}}}\times\left(\boldsymbol{\nabla}_{{\scriptscriptstyle \boldsymbol{R}_{j}}}\times\mathbf{G}\right),\\
\Big(\boldsymbol{\gamma}_{ij}^{(T,2s)}\Big)^{[1]}= & \,-\cfrac{28\pi\eta b^{2}}{3}\,\gamma^{T}\,\mathcal{F}_{i}^{0}\mathcal{F}_{j}^{1}\,\boldsymbol{\nabla}_{{\scriptscriptstyle \boldsymbol{R}_{j}}}\mathbf{G},\quad\qquad\qquad\qquad\quad & \Big(\boldsymbol{\gamma}_{ij}^{(R,2s)}\Big)^{[1]}=\,-\frac{28\pi\eta b^{2}}{6}\,\gamma^{R}\,\boldsymbol{\nabla}_{{\scriptscriptstyle \boldsymbol{R}_{i}}}\times\left(\boldsymbol{\nabla}_{{\scriptscriptstyle \boldsymbol{R}_{j}}}\mathbf{G}\right),\\
\Big(\boldsymbol{\gamma}_{ij}^{(T,3a)}\Big)^{[1]}= & \,-\cfrac{13\pi\eta b^{3}}{9}\,\gamma^{T}\,\boldsymbol{\nabla}_{{\scriptscriptstyle \boldsymbol{R}_{j}}}(\boldsymbol{\nabla}_{{\scriptscriptstyle \boldsymbol{R}_{j}}}\times\mathbf{G}),\qquad\quad\qquad\qquad & \Big(\boldsymbol{\gamma}_{ij}^{(R,3a)}\Big)^{[1]}=\,-\frac{13\pi\eta b^{3}}{18}\,\gamma^{R}\boldsymbol{\nabla}_{{\scriptscriptstyle \boldsymbol{R}_{i}}}\times\left(\boldsymbol{\nabla}_{{\scriptscriptstyle \boldsymbol{R}_{j}}}(\boldsymbol{\nabla}_{{\scriptscriptstyle \boldsymbol{R}_{j}}}\times\mathbf{G})\right),\\
\Big(\boldsymbol{\gamma}_{ij}^{(T,3t)}\Big)^{[1]}\,= & \cfrac{4\pi\eta b^{3}}{5}\,\gamma^{T}\,\boldsymbol{\nabla}_{{\scriptscriptstyle \boldsymbol{R}_{j}}}^{2}\mathbf{G},\qquad\qquad\quad\,\,\,\qquad\qquad & \Big(\boldsymbol{\gamma}_{ij}^{(R,3t)}\Big)^{[1]}\,=\,0,\\
\Big(\boldsymbol{\gamma}_{ij}^{(T,4a)}\Big)^{[1]}= & \cfrac{121\pi\eta b^{4}}{10}\,\gamma^{T}\boldsymbol{\nabla}_{{\scriptscriptstyle \boldsymbol{R}_{j}}}\boldsymbol{\nabla}_{{\scriptscriptstyle \boldsymbol{R}_{j}}}(\boldsymbol{\nabla}_{{\scriptscriptstyle \boldsymbol{R}_{j}}}\times\mathbf{G}),\quad\qquad & \Big(\boldsymbol{\gamma}_{ij}^{(R,4a)}\Big)^{[1]}=\frac{121\pi\eta b^{4}}{20}\,\gamma^{R}\boldsymbol{\nabla}_{{\scriptscriptstyle \boldsymbol{R}_{i}}}\times\left(\boldsymbol{\nabla}_{{\scriptscriptstyle \boldsymbol{R}_{j}}}\boldsymbol{\nabla}_{{\scriptscriptstyle \boldsymbol{R}_{j}}}(\boldsymbol{\nabla}_{{\scriptscriptstyle \boldsymbol{R}_{j}}}\times\mathbf{G})\right).
\end{eqnarray*}
Here $\gamma^{T}=6\pi\eta b$, $\gamma^{R}=8\pi\eta b{}^{3}$. The
friction tensor corresponding to $l\sigma=2t$ are obtained in terms
of the pressure Green's function as
\begin{eqnarray}
\boldsymbol{\gamma}_{ij}^{(2t,\,T)} & = & \mathbf{P}(\boldsymbol{R}_{i},\boldsymbol{R}_{j}),\quad\boldsymbol{\gamma}_{ij}^{(2t,\,ls)}=\tfrac{l(l-1)}{2(2l-1)}\,\boldsymbol{\nabla}_{{\scriptscriptstyle \boldsymbol{R}_{j}}}^{(l)}\mathbf{P}(\boldsymbol{R}_{i},\boldsymbol{R}_{j}),\quad\boldsymbol{\gamma}_{ij}^{(2t,\,la)}=0,\quad\boldsymbol{\gamma}_{ij}^{(2t,\,lt)}=0.
\end{eqnarray}
At this level, the hydrodynamic interactions are pairwise-additive
and add to the one-body solutions for the force and torque obtained
at the zeroth iteration. The resulting force and torque on the active
colloids are \emph{geometrically} isotropic and \emph{hydrodynamically}
anisotropic due to the active slip. An intuitive understanding of
the same can be gained from Fig.(\ref{fig:orientation-dependence}).
The first panel shows the fluid flow around a passive particle which
is determined solely by the sum of the body forces. Changing the particle
orientation does not change the flow and, therefore, produces no change
to the forces, shown by the white arrows, on the two test particles.
The second panel plots the flow around an active particle where the
slip contains the modes $l\sigma=2s$ and $l\sigma=3t$. In the third
panel, the active particle is rotated clockwise by $\pi/2$, without
any change in its position. The forces are now different, even though
there has been no changes in relative positions. Similar considerations
apply for the torque as the reader can easily verify.

The above analysis is then used to to obtain explicit forms of the
mobility matrices and the propulsion tensors. Their leading order
forms, for $i\neq j$, are
\begin{eqnarray*}
\bm{\mu}_{ij}^{TT}\quad\,= & \,\,\mathcal{F}_{i}^{0}\mathcal{F}_{j}^{0}\mathbf{G},\qquad\qquad\qquad\qquad\qquad\qquad & \bm{\mu}_{ij}^{RT}\quad\,=\frac{1}{2}\,\boldsymbol{\nabla}_{{\scriptscriptstyle \boldsymbol{R}_{i}}}\times\mathbf{G},\\
\bm{\mu}_{ij}^{TR}\quad\,= & \,\,\,\,\cfrac{1}{2}\,\boldsymbol{\nabla}_{{\scriptscriptstyle \boldsymbol{R}_{j}}}\times\mathbf{G},\qquad\qquad\qquad\qquad\quad\qquad & \bm{\mu}_{ij}^{RR}\quad\,=\frac{1}{4}\,\boldsymbol{\nabla}_{{\scriptscriptstyle \boldsymbol{R}_{i}}}\times\left(\boldsymbol{\nabla}_{{\scriptscriptstyle \boldsymbol{R}_{j}}}\times\mathbf{G}\right),\\
\bm{\pi}_{ij}^{(T,\,2s)}= & \cfrac{28\pi\eta b^{2}}{3}\,\mathcal{F}_{i}^{0}\mathcal{F}_{j}^{1}\,\boldsymbol{\nabla}_{{\scriptscriptstyle \boldsymbol{R}_{j}}}\mathbf{G},\,\,\,\,\,\qquad\qquad\quad & \bm{\pi}_{ij}^{(R,\,2s)}=\,\frac{28\pi\eta b^{2}}{6}\,\boldsymbol{\nabla}_{{\scriptscriptstyle \boldsymbol{R}_{i}}}\times\left(\boldsymbol{\nabla}_{{\scriptscriptstyle \boldsymbol{R}_{j}}}\mathbf{G}\right),\\
\bm{\pi}_{ij}^{(T,\,3a)}= & \cfrac{13\pi\eta b^{3}}{9}\,\boldsymbol{\nabla}_{{\scriptscriptstyle \boldsymbol{R}_{j}}}(\boldsymbol{\nabla}_{{\scriptscriptstyle \boldsymbol{R}_{j}}}\times\mathbf{G}),\quad\qquad\qquad\,\, & \bm{\pi}_{ij}^{(R,\,3a)}=\,\frac{13\pi\eta b^{3}}{18}\,\boldsymbol{\nabla}_{{\scriptscriptstyle \boldsymbol{R}_{i}}}\times\left(\boldsymbol{\nabla}_{{\scriptscriptstyle \boldsymbol{R}_{j}}}(\boldsymbol{\nabla}_{{\scriptscriptstyle \boldsymbol{R}_{j}}}\times\mathbf{G})\right),\\
\bm{\pi}_{ij}^{(T,\,3t)}= & -\cfrac{4\pi\eta b^{3}}{5}\,\boldsymbol{\nabla}_{{\scriptscriptstyle \boldsymbol{R}_{j}}}^{2}\mathbf{G},\quad\,\qquad\quad\,\,\,\qquad\qquad & \bm{\pi}_{ij}^{(R,\,3t)}=0,\\
\bm{\pi}_{ij}^{(T,\,4a)}= & \,-\cfrac{121\pi\eta b^{4}}{10}\,\boldsymbol{\nabla}_{{\scriptscriptstyle \boldsymbol{R}_{j}}}\boldsymbol{\nabla}_{{\scriptscriptstyle \boldsymbol{R}_{j}}}(\boldsymbol{\nabla}_{{\scriptscriptstyle \boldsymbol{R}_{j}}}\times\mathbf{G}),\qquad & \bm{\pi}_{ij}^{(R,\,4a)}=-\frac{121\pi\eta b^{4}}{20}\,\boldsymbol{\nabla}_{{\scriptscriptstyle \boldsymbol{R}_{i}}}\times\left(\boldsymbol{\nabla}_{{\scriptscriptstyle \boldsymbol{R}_{j}}}\boldsymbol{\nabla}_{{\scriptscriptstyle \boldsymbol{R}_{j}}}(\boldsymbol{\nabla}_{{\scriptscriptstyle \boldsymbol{R}_{j}}}\times\mathbf{G})\right).
\end{eqnarray*}

\section{Boundary integrals for fluid velocity and pressure\label{appendix:fluid-vel-pressure}}

Here, we provide explicit expression for the boundary integrals in
Eq.(\ref{eq:bi-v-p}). These are given in terms of a Green's function
of Stokes equation, the corresponding pressure vector, and their derivatives
\cite{singh2015many}\begin{subequations}
\begin{eqnarray}
\boldsymbol{G}_{j}^{(l)}(\boldsymbol{r},\,\boldsymbol{R}_{j})= & \,b^{l-1}\mathcal{F}_{j}^{l-1}\mathbf{\bm{\nabla}}_{{\scriptscriptstyle \boldsymbol{R}_{j}}}^{(l-1)}\mathbf{G}(\boldsymbol{r},\boldsymbol{R}_{j}),\qquad & \boldsymbol{K}_{j}^{(l)}(\boldsymbol{r},\,\boldsymbol{R}_{j})=\frac{4\pi b^{l}}{(l-2)!(2l-1)!!}\mathcal{F}_{j}^{l-1}\mathbf{\bm{\nabla}}_{{\scriptscriptstyle \boldsymbol{R}_{j}}}^{(l-2)}\mathbf{K}(\boldsymbol{r},\boldsymbol{R}_{j}),\\
\boldsymbol{P}_{j}^{(l)}(\boldsymbol{r},\,\boldsymbol{R}_{j})= & \,b^{l-1}\mathbf{\bm{\nabla}}_{{\scriptscriptstyle \boldsymbol{R}_{j}}}^{(l-1)}\mathbf{P}(\boldsymbol{r},\,\boldsymbol{R}_{j}),\qquad\qquad & \boldsymbol{Q}_{j}^{(l)}(\boldsymbol{r},\,\boldsymbol{R}_{j})=\frac{2\pi b^{l}}{(l-2)!(2l-1)!!}\mathbf{\bm{\nabla}}_{{\scriptscriptstyle \boldsymbol{R}_{j}}}^{(l-1)}\mathbf{P}(\boldsymbol{r},\,\boldsymbol{R}_{j}).
\end{eqnarray}
\end{subequations}We use the above expression, and the generalized
friction tensors, to obtain explicit expression of the fluid flow
and pressure. Their explicit expressions, in terms of knowns, are
\begin{alignat}{1}
\boldsymbol{u}(\boldsymbol{r}) & =-\boldsymbol{G}_{j}^{(1s)}(\boldsymbol{r},\boldsymbol{R}_{i})\cdot\mathbf{F}_{j}^{{H}}-\boldsymbol{G}_{j}^{(2a)}(\boldsymbol{r},\,\boldsymbol{R}_{j})\cdot\mathbf{T}_{j}^{{H}}-\sum_{l\sigma=2s}^{\infty}\Big(\sum_{l'\sigma'=2s}^{\infty}\boldsymbol{G}_{i}^{(l'\sigma')}(\boldsymbol{r},\,\boldsymbol{R}_{i})\cdot\boldsymbol{\gamma}_{ij}^{(l'\sigma',\,l\sigma)}-\eta\,\boldsymbol{K}_{j}^{(l\sigma)}(\boldsymbol{r},\,\boldsymbol{R}_{j})\Big)\cdot\mathbf{V}_{j}^{(l\sigma)}\nonumber \\
p(\boldsymbol{r}) & =-\boldsymbol{P}_{i}^{(1)}(\boldsymbol{r},\boldsymbol{R}_{i})\cdot\mathbf{F}_{i}^{{H}}-\sum_{l=2}^{\infty}\Big(\boldsymbol{P}_{i}^{(l's)}(\boldsymbol{r},\boldsymbol{R}_{i})\cdot\boldsymbol{\gamma}_{ij}^{(l's,\,ls)}-\boldsymbol{Q}_{j}^{(ls)}(\boldsymbol{r},\boldsymbol{R}_{j})\Big)\cdot\mathbf{V}_{j}^{(ls)}.
\end{alignat}
These are then used to obtain the irreducible expressions of the fluid
flow and the pressure in Eq.(\ref{eq:p-v-knowns}). The tensors relating
the irreducible slip modes to fluid velocity and pressure are
\begin{gather}
\boldsymbol{\Pi}_{j}^{(l\sigma)}=-\Big(\boldsymbol{G}_{i}^{(l'\sigma')}(\boldsymbol{r},\,\boldsymbol{R}_{i})\cdot\boldsymbol{\gamma}_{ij}^{(l'\sigma',\,l\sigma)}-\,\boldsymbol{K}_{j}^{(l\sigma)}(\boldsymbol{r},\,\boldsymbol{R}_{j})\Big),\quad\boldsymbol{\Lambda}_{j}^{(ls)}=-\Big(\boldsymbol{P}_{i}^{(l's)}(\boldsymbol{r},\boldsymbol{R}_{i})\cdot\boldsymbol{\gamma}_{ij}^{(l's,\,ls)}-\boldsymbol{Q}_{j}^{(ls)}(\boldsymbol{r},\boldsymbol{R}_{j})\Big).
\end{gather}
In the above expressions for the fluid flow and pressure, we have
used the definition of projection operators Eq.(\ref{eq:project-vlsigma})
to define the irreducible parts of the boundary integrals as: $\boldsymbol{G}_{i}^{(l\sigma)}=\mathbf{P}^{(l\sigma)}\cdot\boldsymbol{G}_{i}^{(l)}$,
$\boldsymbol{K}_{i}^{(l\sigma)}=\mathbf{P}^{(l\sigma)}\cdot\boldsymbol{K}_{i}^{(l)}$,
etc.

\begin{table*}
\renewcommand{\arraystretch}{2} \centering

\begin{tabular}{|>{\centering}b{1.5cm}|>{\centering}p{5cm}|>{\centering}p{4.8cm}|>{\centering}p{5cm}|}
\hline 
$l\sigma$ & $v_{\rho}$ & $v_{\theta}$ & $v_{\varphi}$\tabularnewline
\hline 
\hline 
$2s$ & $V_{0}^{(2s)}\big(\tfrac{2}{3}-\sin^{2}\theta\big)$ & $-\tfrac{1}{2}V_{0}^{(2s)}\sin2\theta$ & 0\tabularnewline
\hline 
$3s$ & $V_{0}^{(3s)}\cos\theta\left(\cos^{2}\theta-\tfrac{1}{15}\right)$ & $-V_{0}^{(3s)}\sin\theta\left(\cos^{2}\theta-\tfrac{1}{5}\right)$ & 0\tabularnewline
\hline 
$3a$ & 0 & 0 & $\frac{1}{18}V_{0}^{(3a)}\sin2\theta$\tabularnewline
\hline 
$3t$ & $\tfrac{1}{45}V_{0}^{(3t)}\cos\theta$ & $\tfrac{1}{45}V_{0}^{(3t)}\sin\theta$ & 0\tabularnewline
\hline 
$4a$ & 0 & 0 & $\tfrac{1}{60}V_{0}^{(4a)}\sin\theta\left(\cos^{2}\theta-\tfrac{1}{5}\right)$\tabularnewline
\hline 
\end{tabular}\caption{Active slip velocity $\boldsymbol{v}^{\mathcal{A}}$ in terms of spherical
polar coordinates $(\boldsymbol{\hat{\rho}},\,\boldsymbol{\hat{\theta}},\,\boldsymbol{\hat{\varphi}})$,
for leading coefficients of polar, apolar and chiral symmetry (Eq.(\ref{eq:-slip-truncate})).
We have used orientation $\boldsymbol{p}$ of the colloids for uniaxial
parametrization of the coefficients of the slip expansion (Eq.(\ref{eq:uniaxial-parametrization})).
Without any loss of generality, we choose $\boldsymbol{p}$ to be
along $\hat{\boldsymbol{z}}$-axis, such that $\boldsymbol{p}=\cos\theta\,\hat{\boldsymbol{\rho}}-\sin\theta\,\hat{\boldsymbol{\theta}}$.
\label{tab:Slip-velocity-on} }
\end{table*}
\begin{table}[H]
\renewcommand{\arraystretch}{2}\centering

\begin{tabular}{|>{\centering}m{2.8cm}|>{\centering}p{14cm}|}
\hline 
 & Green's function\tabularnewline
\hline 
\hline 
Unbounded fluid & $G_{\alpha\beta}^{\text{o}}(\boldsymbol{R}_{i}-\boldsymbol{R}_{j})=\frac{1}{8\pi\eta}\left(\nabla^{2}\delta_{\alpha\beta}-\nabla_{\alpha}\nabla_{\beta}\right)\boldsymbol{r}_{ij}$.\tabularnewline
\hline 
Plane interface (fluid-gas) & $G_{\alpha\beta}^{\text{i}}(\boldsymbol{R}_{i},\,\boldsymbol{R}_{j})=G_{\alpha\beta}^{\text{o}}(\boldsymbol{r}_{ij})+(\delta_{\beta\rho}\delta_{\rho\gamma}-\delta_{\beta3}\delta_{3\gamma})G_{\alpha\gamma}^{\text{o}}(\boldsymbol{r}_{ij}^{*})$.\tabularnewline
\hline 
Plane interface (fluid-fluid) & $G_{\alpha\beta}^{\text{f}}(\boldsymbol{R}_{i},\,\boldsymbol{R}_{j})=G_{\alpha\beta}^{\text{o}}(\boldsymbol{r}_{ij})+\mathcal{M}_{\beta\gamma}^{f}G_{\alpha\gamma}^{\text{o}}(\boldsymbol{r}_{ij}^{*})-2h\tfrac{\lambda}{1+\lambda}\nabla_{{\scriptscriptstyle \boldsymbol{r}_{\gamma}^{*}}}G_{\alpha3}^{\text{o}}(\boldsymbol{r}_{ij}^{*})\mathcal{M}_{\beta\gamma}+h^{2}\tfrac{\lambda}{1+\lambda}\nabla_{{\scriptscriptstyle \boldsymbol{r}^{*}}}^{2}G_{\alpha\gamma}^{\text{o}}(\boldsymbol{r}_{ij}^{*})\mathcal{M}_{\beta\gamma}.$\tabularnewline
\hline 
Plane no-slip wall & $G_{\alpha\beta}^{\text{w}}(\boldsymbol{R}_{i},\,\boldsymbol{R}_{j})=G_{\alpha\beta}^{\text{o}}(\boldsymbol{r}_{ij})-G_{\alpha\beta}^{\text{o}}(\boldsymbol{r}_{ij}^{*})-2h\nabla_{{\scriptscriptstyle \boldsymbol{r}_{\gamma}^{*}}}G_{\alpha3}^{\text{o}}(\boldsymbol{r}_{ij}^{*})\mathcal{M}_{\beta\gamma}+h^{2}\nabla_{{\scriptscriptstyle \boldsymbol{r}^{*}}}^{2}G_{\alpha\gamma}^{\text{o}}(\boldsymbol{r}_{ij}^{*})\mathcal{M}_{\beta\gamma}.$\tabularnewline
\hline 
\end{tabular}\caption{The Green's functions of Stokes equation for a system of two colloids
at $\boldsymbol{R}_{i}$ and $\boldsymbol{R}_{j}$ respectively \cite{blake1971c,blake1975,aderogba1978action}.
$\boldsymbol{r}_{ij}=\mathbf{\boldsymbol{R}}_{i}-\mathbf{\boldsymbol{R}}_{j}$,
and $\boldsymbol{r}_{ij}^{*}=\mathbf{\boldsymbol{R}}_{i}-\mathbf{\boldsymbol{R}}_{j}^{*}$,
where $\boldsymbol{R}_{j}^{*}=\boldsymbol{\mathcal{M}}\cdot\boldsymbol{R}$
is the image of the $j$th colloid at a distance $h$ from the interface/wall
at $z=0$. $\boldsymbol{\mathcal{M}}=\boldsymbol{I}-2\mathbf{\hat{z}}\mathbf{\hat{z}}$,
$\lambda=\eta_{2}/\eta_{1}$, and $\mathcal{M}_{\beta\gamma}^{f}=\left(\tfrac{1-\lambda}{1+\lambda}\delta_{\beta\rho}\delta_{\rho\gamma}-\delta_{\beta3}\delta_{3\gamma}\right)$,
with $\rho$ taking values $1,\,2$ in the plane of the interface.
\label{tab:geometries-G}}
\end{table}
 \clearpage

\section{Symbols and notations}

A list of symbols and notations used in the paper is given below.
The first column describes quantities which have particle indices
while those in the second column do not have particle indices. The
list has been ordered based on the first appearance of these symbols
in the main text. We use abbreviate rigid body motion as RBM and boundary
integral equation as BIE.

\begin{minipage}[t]{0.44\textwidth}%
\begin{eqnarray*}
\mathbf{F}_{i}^{H} & \quad & \text{hydrodynamic forces on the \ensuremath{i}-th colloid}\\
\mathbf{F}_{i}^{P} & \quad & \text{body force on the \ensuremath{i}-th colloid}\\
\hat{\mathbf{F}}_{i} & \quad & \text{Brownian force on the \ensuremath{i}-th colloid}\\
\mathbf{T}_{i}^{H} & \quad & \text{hydrodynamic torque on the \ensuremath{i}-th colloid}\\
\mathbf{T}_{i}^{P} & \quad & \text{body torque on the \ensuremath{i}-th colloid}\\
\hat{\mathbf{T}}_{i} & \quad & \text{Brownian torque on the \ensuremath{i}-th colloid}\\
\mathbf{V}_{i} & \quad & \text{translational velocity of the \ensuremath{i}-th colloid}\\
\mathbf{\Omega}_{i} & \quad & \text{angular velocity of the \ensuremath{i}-th colloid}\\
\mathbf{V}_{i}^{\mathcal{A}} & \quad & \text{active translational velocity of the \ensuremath{i}-th colloid}\\
\mathbf{\Omega}_{i}^{\mathcal{A}} & \quad & \text{active angular velocity of the \ensuremath{i}-th colloid}\\
\boldsymbol{R}_{i} & \quad & \text{coordinate of the centre of the \ensuremath{i}-th colloid}\\
\boldsymbol{p}_{i} & \quad & \text{orientation of the \ensuremath{i}-th colloid}\\
\boldsymbol{v} & \quad & \text{boundary velocity on a colloid}\\
\boldsymbol{v}_{i}^{\mathcal{A}} & \quad & \text{\text{active slip at the surface of the \ensuremath{i}-th colloid}}\\
\boldsymbol{f} & \quad & \text{force per unit area (or the traction)}\\
\boldsymbol{r}_{i} & \quad & \text{a point on the surface of the \ensuremath{i}-th colloid}\\
\boldsymbol{r}_{ij} & \quad & \boldsymbol{R}_{i}-\boldsymbol{R}_{i},\text{ separation between colloids}\\
S_{i} & \quad & \text{surface of the \ensuremath{i}-th colloid}\\
\boldsymbol{\rho}_{i} & \quad & \text{radius vector of the \ensuremath{i}-th colloid}\\
\mathbf{Y}^{(l)} & \quad & l\text{-th tensorial spherical harmonics}\\
\mathbf{V}^{(l)} & \quad & l\text{-th coefficient of slip expansion}\\
\mathbf{F}^{(l)} & \quad & l\text{-th coefficient of traction expansion}\\
\mathbf{P}^{(l\sigma)} & \quad & \text{Projection operator }\\
\mathbf{F}_{i}^{(ls)} & \quad & \text{symmetric irreducible part of }\mathbf{F}_{i}^{(l)}\\
\mathbf{F}_{i}^{(la)} & \quad & \text{antisymmetric part of }\mathbf{F}_{i}^{(l)}\\
\mathbf{F}_{i}^{(lt)} & \quad & \text{trace of }\mathbf{F}_{i}^{(l)}\\
\boldsymbol{\mu}_{ij}^{\alpha\beta} & \quad & \text{mobility matrix for }\alpha,\beta=T,R\\
\boldsymbol{\pi}_{ij}^{(\alpha,\,l\sigma)} & \quad & \text{propulsion tensors}\\
\boldsymbol{\gamma}_{ij}^{(l'\sigma',\,l\sigma)} & \quad & \text{generalized friction tensors}\\
\boldsymbol{G}_{i}^{(l)} & \quad & \text{single-layer integral for fluid flow}\\
\boldsymbol{K}_{i}^{(l)} & \quad & \text{double-layer integral for fluid flow}\\
\boldsymbol{P}_{i}^{(l)} & \quad & \text{single-layer integral for fluid pressure}\\
\boldsymbol{Q}_{i}^{(l)} & \quad & \text{double-layer integral for fluid pressure}\\
\boldsymbol{G}_{ij}^{(l,\,l')} & \quad & \text{single-layer matrix elements}\\
\boldsymbol{K}_{ij}^{(l,\,l')} & \quad & \text{double-layer matrix elements}\\
\mathcal{F}_{i}^{l} & \quad & \text{operator encoding finite size of colloids}
\end{eqnarray*}
\end{minipage}$\qquad\quad$%
\begin{minipage}[t]{0.44\textwidth}%
\begin{eqnarray*}
b & \quad & \text{radius of the colloid}\\
N & \quad & \text{number of colloids}\\
\eta & \quad & \text{fluid viscosity}\\
\boldsymbol{\sigma} & \quad & \text{fluid stress}\\
\boldsymbol{u} & \quad & \text{fluid velocity}\\
p & \quad & \text{fluid pressure}\\
\boldsymbol{u}^{\infty} & \quad & \text{externally imposed flow}\\
\mathbf{G} & \quad & \text{a Green's function of Stokes equation}\\
\mathbf{K} & \quad & \text{a Stress function of Stokes equation}\\
\boldsymbol{G} & \quad & \text{single-layer operator of the BIE}\\
\boldsymbol{K} & \quad & \text{double-layer operator of the BIE}\\
\boldsymbol{\xi} & \quad & \text{thermal force acting on the fluid }\\
V & \quad & \text{volume of the fluid}\\
\phi & \quad & \text{volume fraction}\\
\boldsymbol{\mathcal{\dot{E}}} & \quad & \text{power dissipation in the fluid}\\
\mathcal{A}_{T} & \quad & \text{translational Activity number}\\
\mathcal{A}_{R} & \quad & \text{rotational Activity number}\\
\mathcal{B}_{T} & \quad & \text{translational Brown number}\\
\mathcal{B}_{R} & \quad & \text{rotational Brown number}\\
\boldsymbol{\Psi} & \quad & \text{distribution function for colloids}\\
c^{(n)} & \quad & \text{\textit{n}-body density}\\
\boldsymbol{\Sigma}^{H} & \quad & \text{hydrodynamic suspension stress}\\
\boldsymbol{\Sigma}^{P} & \quad & \text{particle contribution to the stress}\\
\boldsymbol{E} & \quad & \text{macroscopic strain rate}\\
\boldsymbol{\Delta}^{(l)} & \quad & \mathbf{F}_{i}^{(ls)}=\boldsymbol{\Delta}^{(l)}\cdot\mathbf{F}_{i}^{(l)}\\
\boldsymbol{\varepsilon} & \quad & \text{the Levi-Civita tensor}\\
\boldsymbol{\delta} & \quad & \text{the Kronecker delta}\\
\boldsymbol{I} & \quad & \text{identity tensor}\\
v_{s} & \quad & \text{self-propulsion speed of the colloid}\\
\omega_{s} & \quad & \text{self-rotation speed of the colloid}\\
U & \quad & \text{external potential}\\
\epsilon & \quad & \text{strength of the WCA potential}\\
k & \quad & \text{stiffness of the harmonic trap trap}\\
k^{c} &  & \text{strength of the spherical confinement}\\
\tau & \quad & \text{rotational time scale }V_{0}^{(4a)}/b\\
\tau_{r} & \quad & \text{rotational time scale in trap }8\pi\eta b/k\\
k_{B} & \quad & \text{Boltzmann constant}\\
T & \quad & \text{temperature}
\end{eqnarray*}
\end{minipage}

\end{document}